\documentclass[aps,prb, amsfonts, amssymb, amsmath, superscriptaddress, notitlepage,twocolumn,10pt,nobalancelastpage,longbibliography]{revtex4-2}
\usepackage{graphicx} 
\usepackage{hyperref}
\usepackage{bm}
\hypersetup{ 
    colorlinks=true, 
    linktoc=all,     
    linkcolor=blue,  
    citecolor=blue
} 
\usepackage{mathtools}
\usepackage{amsmath}
\usepackage{natbib}

\newcommand{\bk}{\mathbf{k}}
\newcommand{\bK}{\mathbf{K}}
\newcommand{\bq}{\mathbf{q}}

\newcommand{\bs}{\mathbf{s}}
\newcommand{\bS}{\mathbf{S}}

\newcommand{\ba}{\mathbf{a}}
\newcommand{\bA}{\mathbf{A}}

\newcommand{\nn}{\nonumber}

\begin{document}

\title{Instabilities of the Fractionalized Dirac Semimetal in the Kitaev-Kondo Model}

\author{Jennifer Lin}
\affiliation{London Centre for Nanotechnology, University College London, Gordon St., London, WC1H 0AH, United Kingdom}
\author{Frank Kr\"uger}
\affiliation{London Centre for Nanotechnology, University College London, Gordon St., London, WC1H 0AH, United Kingdom}
\affiliation{ISIS Facility, Rutherford Appleton Laboratory, Chilton, Didcot, Oxfordshire OX11 0QX, United Kingdom}

\begin{abstract}
We study a honeycomb Kondo lattice model in which Dirac conduction electrons are coupled to a spin-1/2 Kitaev quantum spin liquid. For weak Kondo coupling, 
the spins fractionalize into Majorana fermions comprising a gapless Dirac mode and three gapped visons. 
In second order perturbation theory, the Kondo coupling gives rise to local Hubbard repulsions and  spin–spin interactions between conduction electrons, 
as well as a vertex coupling electrons to gapless Majorana fermions. 
We analyze the resulting low-energy field theory using a perturbative renormalization group (RG) scheme,  accounting for additional density–density interactions 
generated under RG. At criticality, electrons decouple from Majorana fermions but all three electron interactions acquire positive values.  An analysis of susceptibility 
exponents reveals that the fractionalized Fermi liquid becomes unstable towards antiferromagnetic order and that 
superconductivity is disfavored.
\end{abstract}

\maketitle

\section{Introduction}
\label{sec.intro}

Quantum spin liquids (QSLs) represent one of the most striking manifestations of many-body quantum physics. In
these states, local magnetic moments fail to order even at zero temperature, remaining in a highly entangled and
dynamic quantum state. This phenomenon arises from geometric frustration and strong quantum fluctuations,
producing a “liquid-like” magnetic phase in which fractionalized excitations and gauge fields naturally
emerge \cite{Anderson73,Balents10}.

In fractionalized Fermi liquids, dubbed as FL* phases, conduction electrons coexist with a QSL 
background. Such exotic states were originally proposed as phases of Kondo lattice models \cite{Senthil+03,Senthil+04}.
While for strong Kondo coupling the electrons that initially form the local moment spins hybridize with the 
conduction electrons, resulting in a heavy Fermi liquid with a large Fermi surface, for sufficiently weak Kondo coupling 
the tendency of spin fractionalization dominates over the Kondo screening.  Since the emergent fractionalized 
quasiparticles don't carry electrical charge, the resulting FL* state has a small Fermi surface of the conduction electrons alone, 
leading to an apparent violation of Luttinger's theorem \cite{Hofmann+19}.

More recently, fractionalized Fermi liquids have been proposed to underlie the pseudo-gap regime
of under-doped cuprate superconductors \cite{Qi+10,Moon+11,Mei+12}, motivated by the experimental observation 
of small Fermi pockets and Fermi arcs, and reviving the early proposal by P.W. Anderson that the cuprates 
may be understood as doped quantum spin liquids \cite{Anderson87}. 

However, the physics of the cuprates is extremely rich. It is therefore important to first understand 
fractionalized Fermi liquids and their superconducting pairing instabilities on the level of simple toy models. One such setting 
is a Kondo-lattice model in which tight-binding electrons on the honeycomb lattice are coupled to local moment spins that form a Kitaev 
QSL \cite{Seifert+18,Choi+18}. 

The spin-1/2 Kitaev model on the honeycomb lattice is one of the very rare examples of an exactly solvable QSL model \cite{Kitaev06}. 
The spins are found to fractionalize into a set of Majorana fermions, one spinon mode with a gapless Dirac dispersion 
and three dispersion-less vison modes that encode local excitations of $Z_2$ gauge fluxes. Although the exact solvability is broken 
by the Kondo coupling to conduction electrons, the vison gap guarantees that the QSL and hence the FL* state remain stable agains 
sufficiently weak Kondo coupling. 

Utilizing Majorana-fermion mean-field theory, it was found that for a ferromagnetic Kitaev model a nematic triplet superconductor (SC) forms 
at intermediate Kondo coupling, sandwiched between the FL* state and the heavy fermion liquid, suggesting that the itinerant Majorana fermions 
act as ``pairing glue"  for unconventional superconductivity \cite{Seifert+18}. The extent of the SC region was found to crucially depend 
upon the conduction electron filling, being maximum for fillings close to the van-Hove singularity and shrinking almost to zero at half filling 
where the Fermi level is located at the Dirac points of the conduction electron spectrum. An alternative mean-field treatment based on 
Abrikosov fermions found a first order instability of the FL* phase towards the formation of a ferromagnetic topological 
superconductor \cite{Choi+18}.

 In a very recent investigation \cite{Bunney+25} the local moment degrees of freedom were integrated out, resulting in a Hubbard repulsion and 
spin-spin interaction between the conduction electrons to second order in the Kondo coupling. The induced electronic spin-spin interaction has the
same sign and bond-directional dependence as the Kitaev exchange. 
 The resulting interacting electron problem was then analyzed using functional renormalization group (fRG). While near the van-Hove filling 
 the FL* was found to become unstable towards the formation of a spin-density wave (SDW), at smaller fillings a transition from the FL* state to 
 a superconductor was found, in the antiferromagnetic case with chiral $d+id$ symmetry, in the ferromagnetic case of spin-triplet $p$-wave type.  
 Unfortunately, the momentum resolution in the fRG was insufficient to investigate the behavior for electron fillings 
 close to the Dirac point. 
  
 In this paper we use a complementary method to investigate the instabilities of the fractionalized half-filled Dirac semimetal of the 
 Kitaev-Kondo model.  Treating the Kondo coupling in second-order perturbation theory,  
 we derive an effective low-energy continuum field theory of the gapless Dirac Majorana fermions coupled to Dirac conduction electrons. 
 It contains a Hubbard on-site repulsion and a spin-spin interactions between conduction electrons, as well as a four-fermion vertex that couples 
 Majorana fermions and conduction electrons. We use a perturbative parquet RG analysis \cite{Murray+14,Chubukov+16,Xing+17,Parthenios+23} to first obtain the 
 scale dependence of the various interaction parameters, and as a second step determine the susceptibility exponents of different order 
 parameters from the RG flow of the conjugate fields at the critical fixed point.
  
 Our analysis shows that  irrespective of the sign of the Kitaev coupling and the ratio of Fermi velocities of Majorana fermions and conduction 
 electrons the critical behavior is always controlled by the same fixed point at which the conduction electrons decouple from the Majorana fermions
 and all electron-electron interactions, including an additional density-density repulsion that is generated under the RG, have finite positive values. 
 The renormalization of various fields conjugate to potential order parameters shows that the leading instability is toward an antiferromagnetic SDW state. 
 
 Our paper is organized as follows. In Sec.~\ref{sec.model} we introduce the Hamiltonian of the Kitaev-Kondo model on the honeycomb lattice and 
 the action of the corresponding path-integral over Grassmann fields. By integrating out the gapped vison modes, in Sec.~\ref{sec.low_en}, 
 we derive the effective low-energy continuum field theory. In Sec.~\ref{sec.RG} the RG equations of the rescaled interaction parameters are obtained. With these 
 results, we identify the critical fixed point for the symmetry-breaking instability of the FL$^*$ phase and analyze the RG 
 flow along the critical surface. To understand the type of symmetry breaking, in Sec.~\ref{sec.fieldRG}, we compute the susceptibility exponents for different 
 order parameters from the field scaling exponents at the critical fixed point. Finally, in Sec.~\ref{sec.disc} we summarize and discuss our results.

\begin{figure}[t!]
 \includegraphics[width=0.85\linewidth]{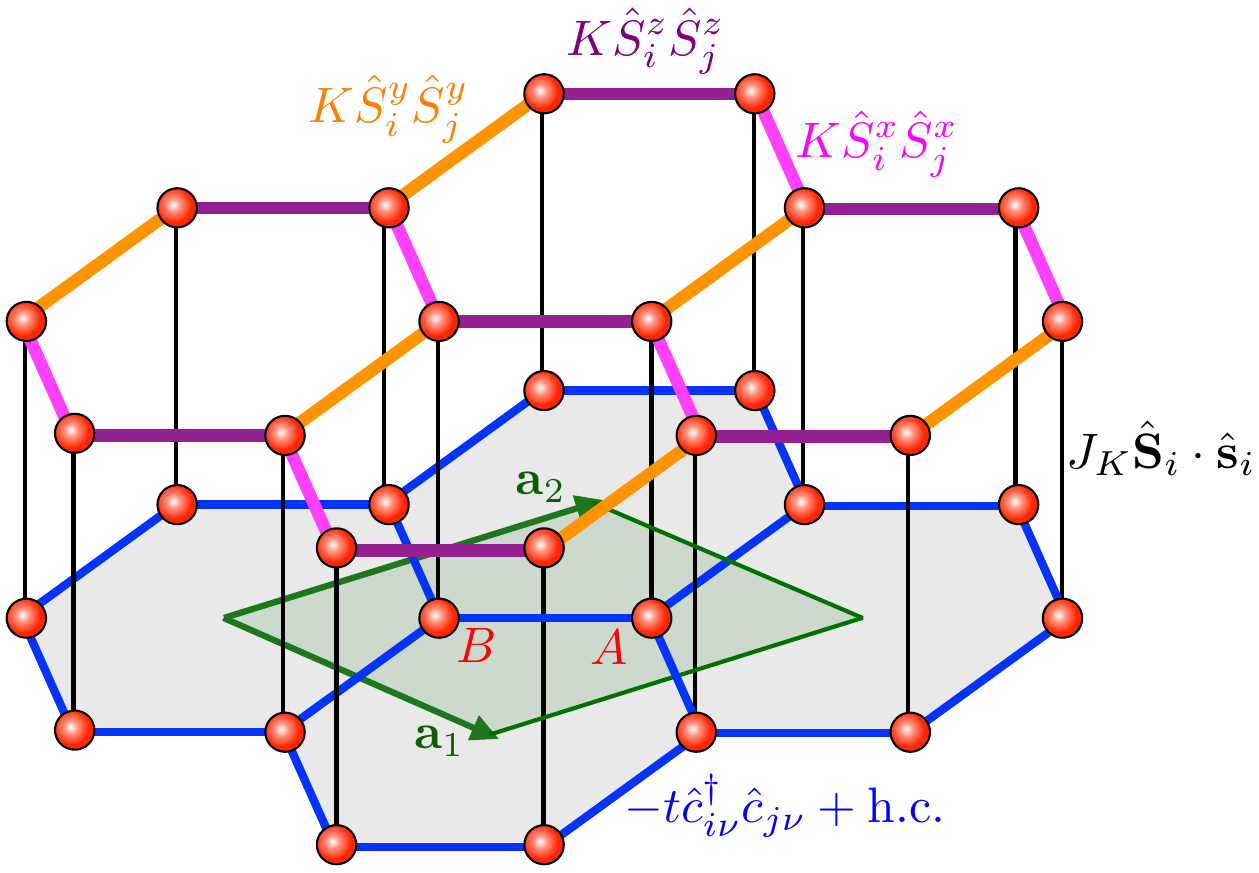}
 \caption{Illustration of the Kitaev-Kondo model on a honeycomb lattice. The top layer illustrates the $S=1/2$ Kitaev QSL model with  
 bond-directional Ising exchanges $K$. The bottom layer corresponds to tight-binding electrons with hopping amplitudes $t$ between neighboring sites of the 
 honeycomb lattice. The local moments of the Kitaev quantum spin liquid are coupled to the conduction electrons via the conventional Kondo interaction $J_K$. A possible
 unit cell spanned by lattice vectors $\ba_{1,2}=(\frac32,\pm\frac{\sqrt{3}}{2})$ is shown in green.}
\label{figure1}
\end{figure}

\section{Model}
\label{sec.model}

Our starting model is the $S=1/2$ Kitaev model on the honeycomb lattice with Kondo coupling to conduction electrons that could either live on the 
same lattice or an adjacent honeycomb layer, as illustrated in Fig.~\ref{figure1}. The Hamiltonian of this model is given by
\begin{eqnarray}
\hat{\mathcal{H}} & = & K \sum_{\gamma=x,y,z} \sum_{\langle i,j\rangle_\gamma} \hat{S}_i^\gamma \hat{S}_j^\gamma-t \sum_{\langle i,j\rangle}  \sum_{\nu =\uparrow,\downarrow} \left(\hat{c}_{i\nu}^\dagger \hat{c}_{j\nu} +\textrm{h.c.}\right)\nn\\
& & +J_K\sum_i\sum_{\gamma}  \sum_{\nu,\nu'}  \hat{S}_j^\gamma \hat{c}_{i\nu}^\dagger \sigma^\gamma_{\nu,\nu'}\hat{c}_{j\nu'}.
\end{eqnarray}
Here $K$ denotes the Kitaev coupling, which could be antiferromagnetic ($K>0$) or ferromagnetic ($K<0$), $t$ the hopping amplitude of the conduction electrons 
between neighboring lattice sites, and $J_K$ the Kondo coupling between 
local moment spins $\hat{\bS}_i$ and conduction electron spins $\hat{\bs}_i$ with components $\hat{s}_i^\gamma = \sum_{\nu,\nu'} \hat{c}_{i\nu}^\dagger 
\sigma^\gamma_{\nu,\nu'}\hat{c}_{j\nu'}$ (in units of $\hbar/2$), where $\bm{\sigma}^\gamma$ denote the standard spin Pauli matrices. In the above Hamiltonian, nearest-neighbor bonds are 
denoted by $\langle i,j\rangle$ and distinguished by the subscript $\gamma=x,y,z$ to express the bond-directional Ising exchange of the Kitaev model (see Fig.~\ref{figure1}).

The tight-binding dispersion of the conduction electrons is given by $\epsilon_{\pm}(\bk) = \pm t | \lambda(\bk)|$, where 
\begin{equation}
\lambda(\bk) = \sum_\gamma e^{i\bk\bm{\delta}_\gamma}
\end{equation}
with $\bm{\delta}_x = \ba_1$, $\bm{\delta}_y = \ba_2$, and $\bm{\delta}_z = \bm{0}$. The lattice vectors $\ba_i$ and corresponding unit cell are defined in Fig.~\ref{figure1}. 
In this work we will focus on the case of half filling, where the Fermi level is located at 
the Dirac points $\bK_\pm = \frac{2\pi}{3} (1,\pm 1/\sqrt{3})$. Linearizing the dispersion near the Dirac points, $\bk=\bK_\pm +\bq$, we obtain 
$\epsilon_{\pm}(\bq) = \pm v| \bq |$ with Fermi velocity $v=\frac32 t$.

The Kitaev model is exactly solvable in terms of a set of four Majorana fermions $\hat{\eta}_i$, $\hat{\xi}_i^x$, $\hat{\xi}_i^y$, $\hat{\xi}_i^z$ \cite{Kitaev06}, which satisfy
$\hat{\eta}_i^\dagger = \hat{\eta}_i$, $(\hat{\xi}_i^\gamma)^\dagger = \hat{\xi}_i^\gamma$, and the Clifford algebra $\{\hat{\xi}_i^\gamma,  \hat{\xi}_j^{\gamma'} \} =2 \delta_{ij}\delta_{\gamma\gamma'}$, $\{\hat{\eta}_i,  \hat{\xi}_j^\gamma \} =0$. In terms of the Majorana fermions the spin $1/2$ operators (in units of $\hbar/2$) are expressed as
\begin{equation}
\hat{S}_i^\gamma = i \hat{\eta}_i \hat{\xi}_i^\gamma,
\end{equation}
where the local constraint $\hat{\eta}_i \hat{\xi}_i^x \hat{\xi}_i^y \hat{\xi}_i^z =1$ ensures that the Hilbert space is not artificially enlarged and that the spin-commutation 
relations $[\hat{S}_i^\alpha, \hat{S}_j^\beta]=2\delta_{ij} \epsilon_{\alpha\beta\gamma}\hat{S}_i^\gamma$ follow from the fermion anti-commutation relations.

Although the resulting Hamiltonian is initially quartic in terms of the Majorana fermion operators, it can be solved analytically since the local bond operators 
$\hat{A}_{ij}^\gamma = i \hat{\xi}_i^\gamma  \hat{\xi}_j^{\gamma}$ and corresponding plaquette operators, given by the product of bond operators around each hexagon, commute with the Hamiltonian.
This results in a free-fermion Hamiltonian with Dirac dispersion $\epsilon_0 (\bk) = \pm K  |\lambda(\bk)|$ of the $\hat{\eta}$ Majorana fermions (spinons) and three flat bands 
 $\epsilon_\gamma (\bk) = \pm \Delta$ for the localized $\hat{\xi}^\gamma$ Majorana fermions (visons)  with an energy gap $\Delta/|K| \approx 0.525$ \cite{Saheli+24,Thiagarajan+24,Bunney+25}. 
 Note that by definition, the Majorana fermions are always at half filling.
 
 To study the effects of the Kondo coupling $J_K$ between Kitaev Majorana fermions and Dirac conduction electrons we employ the coherent state, imaginary-time path 
 integral formalisms. After Fourier transform to frequency $k_0$ and spatial momenta $\bk$ the action is given by $\mathcal{S} = \mathcal{S}_0[\bar{\bm{\psi}},\bm{\psi}] +
 \mathcal{S}_0[\bm{\eta}] + \mathcal{S}_0[\bm{\xi}]+ \mathcal{S}_{\textrm{int}}[\bar{\bm{\psi}},\bm{\psi},\bm{\eta},\bm{\xi}]$ with
\begin{eqnarray}
\mathcal{S}_0[\bar{\bm{\psi}},\bm{\psi}] & = & \sum_{\nu={\uparrow,\downarrow}}\int_k \bar{\bm{\psi}}_\nu(k) \left( \begin{array}{cc} -i k_0 & -t\lambda^*(\bk) \\ -t \lambda(\bk) & -i k_0     \end{array}   \right) \bm{\psi}_\nu(k),\quad \\
\mathcal{S}_0[\bm{\eta}] & = &  \int_k \bm{\eta}^\dagger(k) \left( \begin{array}{cc} -i k_0 & -i K \lambda^*(\bk) \\ i K \lambda(\bk) & -i k_0     \end{array}   \right) \bm{\eta}(k),\\
\mathcal{S}_0[\bm{\xi}^\gamma] & = &  \int_k (\bm{\xi}^\gamma)^\dagger(k) \left( \begin{array}{cc} -i k_0 & -i \Delta e^{-i\bk\bm{\delta}_\gamma} \\ i \Delta e^{i\bk\bm{\delta}_\gamma} & -i k_0     \end{array}   \right) \bm{\xi}^\gamma(k),\\
\mathcal{S}_{\textrm{int}} & = & i J_K \sum_{s=A,B}\sum_\gamma \int_{k_1,\ldots,k_4} \delta_{k_1+k_2-k_3+k_4}\nn\\
& & \times  \eta_s(k_1) \xi_s^\gamma(k_2) \bar{\bm{\psi}}_s(k_3) \bm{\sigma}^\gamma \bm{\psi}_s(k_4).
\end{eqnarray}
Here  $k=(k_0,\bk)$ and $\int_k = \int_{-\infty}^\infty \frac{d k_0}{2\pi} \int_{BZ} \frac{d^2\bk}{V_{BZ}}$, for brevity. Note that while the conduction electrons are represented
by independent Grassmann fields $\bar{\psi}_{s\nu}(k)$, $\psi_{s\nu}(k)$, the Majorana fermions are represented by a single Grassmann field, 
and $\eta_s^\dagger(k) = \eta_s(-k)$ and $(\xi_s^\gamma)^\dagger(k) = \xi_s^\gamma(-k)$.

\section{Effective Low-Energy Theory}
\label{sec.low_en}

To derive an effective low-energy field theory,   we use second 
order perturbation theory to obtain the effective interactions $\sim J_K^2$ between conduction
electrons, as well as an interaction between the gapless Majorana fermions and the conduction electrons.

Using that the fermion Green functions are given by the following 2-by-2 matrices in sub-lattice space, 
\begin{eqnarray}
\bm{G}_\nu^\psi (k) & = &  \frac{1}{k_0^2+t^2 |\lambda(\bk)|^2} \left( \begin{array}{cc} i k_0 & -t\lambda^*(\bk) \\ -t \lambda(\bk) & i k_0     \end{array}   \right),\\
\bm{G}^\eta (k) & = &  \frac{1}{k_0^2+K^2 |\lambda(\bk)|^2} \left( \begin{array}{cc} i k_0 & -iK\lambda^*(\bk) \\ iK \lambda(\bk) & i k_0     \end{array}   \right),\\
\bm{G}^{\xi_\gamma} (k) & = &  \frac{1}{k_0^2+\Delta^2} \left( \begin{array}{cc} i k_0 & -i\Delta e^{-i\bk\bm{\delta}_\gamma} \\ i\Delta e^{i\bk\bm{\delta}_\gamma} & i k_0     \end{array}   \right),
\end{eqnarray}
and that the Majorana fermion Green functions satisfy $G^\eta_{s,s'}(k) = -G^\eta_{s',s}(-k)$ and $G^{\xi_\gamma}_{s,s'}(k) = -G^{\xi_\gamma}_{s',s}(-k)$, 
we obtain an on-site Hubbard repulsion $U$ and a nearest-neighbor spin-spin interaction $J$ between conduction electrons,    

\begin{figure}[t!]
 \includegraphics[width=0.85\linewidth]{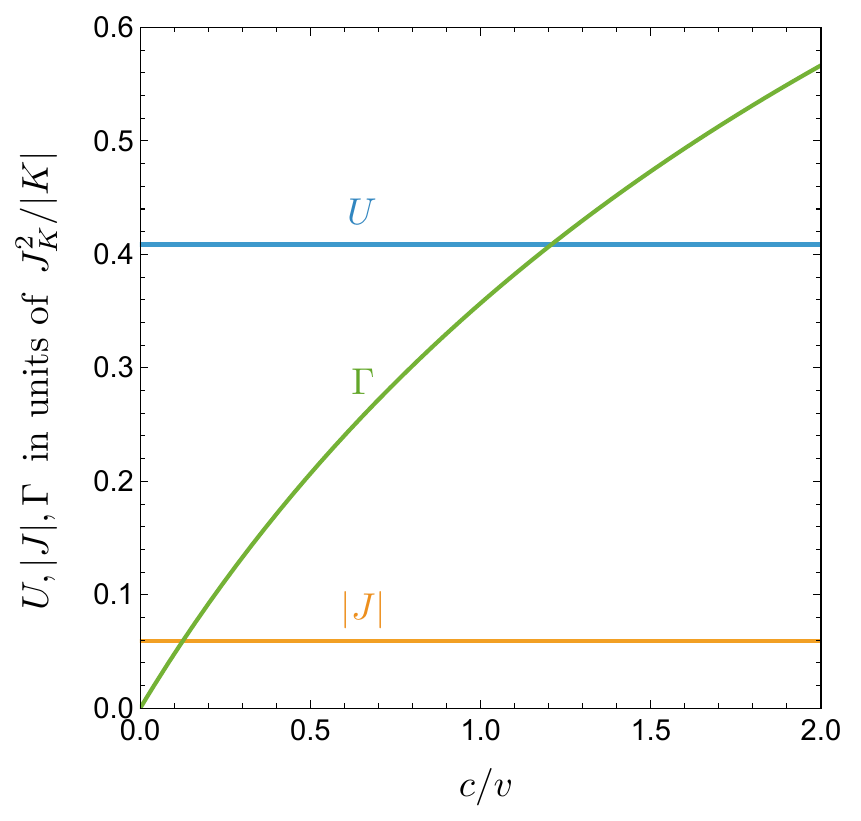}
 \caption{Interaction $U$, $|J|$ and $\Gamma$ in units of $J_K^2/|K|$ and as a function of the ratio of the Fermi velocities $c$ and $v$ of Majorana fermions and 
 Dirac conduction electrons, respectively.}
\label{figure2}
\end{figure}

\begin{eqnarray}
S_U [\bar{\bm{\psi}},\bm{\psi}] & = &  U \sum_{s,\nu}  \int_{k_1,\ldots,k_4} \delta_{k_1-k_2+k_3-k_4}\nn\\
& & \times \bar{\psi}_{s\nu}(k_1)\psi_{s\nu}(k_2)\bar{\psi}_{s\bar{\nu}}(k_3)\psi_{s\bar{\nu}}(k_4),\\
S_J [\bar{\bm{\psi}},\bm{\psi}] & = &  J \sum_{s,\gamma}  \int_{k_1,\ldots,k_4} \delta_{k_1-k_2+k_3-k_4}\nn\\
& & \times \left( \bar{\bm{\psi}}_{s}(k_1)\bm{\sigma}^\gamma \bm{\psi}_{s}(k_2)\right) \left( \bar{\bm{\psi}}_{\bar{s}}(k_3)\bm{\sigma}^\gamma \bm{\psi}_{\bar{s}}(k_4)\right),\quad
\end{eqnarray}
where $\bar{\nu} = \downarrow$ if $\nu=\uparrow$, $\bar{s}=B$ if $s=A$ and vice-versa. 

To obtain the above long-wavelength expression for $\mathcal{S}_J$ we have performed an expansion around the Dirac points. Note that the original 
vertex contains additional exponential factors $e^{i(\bk_3-\bk_4)\bm{\delta}_\gamma}$. This means that the electron spin-spin interaction inherits  the bond-directional 
dependence from the Kitaev model \cite{Bunney+25}. However, taking the continuum limit, the directional dependence is lost at zeroth-order in the gradient expansion. 
We drop the next order terms quadratic in $(\bk_3-\bk_4)\bm{\delta}_\gamma$ since the resulting vertex is irrelevant under the renormalization group. 

The strengths of the interactions $U$ and $J$ are given by the frequency-momentum integrals 
\begin{eqnarray}
U & = & \frac{3 J_K^2}{2}  \int_q \frac{q_0^2}{(q_0^2+K^2|\lambda(\bq)|^2)(q_0^2+\Delta^2)}\approx 0.409 \frac{J_K^2}{|K|},\quad\\
J & = & \frac{J_K^2}{6}  \int_q \frac{K\Delta|\lambda(\bq)|^2}{(q_0^2+K^2|\lambda(\bq)|^2)(q_0^2+\Delta^2)}\approx 0.060 \frac{J_K^2}{K},
\end{eqnarray}
where the numerical results are obtained using $\Delta/|K| \approx 0.525$ \cite{Saheli+24,Thiagarajan+24}.
While $U$ is always positive, the sign of $J$ is set by the sign of the Kitaev coupling $K$, and $|J|/U \approx 0.146$. This is in perfect 
agreement with Ref.~\cite{Bunney+25}, when taking into account their different definition of $U$ and $J$.

In addition to the interactions between conduction electrons, the low-energy theory contains a vertex that couples the conduction electrons to the gapless Majorana fermion, 
\begin{eqnarray}
S_\Gamma [\bar{\bm{\psi}},\bm{\psi},\bm{\eta}] & = &  i\Gamma \sum_{s,\nu} \chi_s   \int_{k_1,\ldots,k_4} \delta_{k_1-k_2+k_3+k_4}\nn\\
& & \times  \eta_s(k_1) \bar{\psi}_{s\nu}(k_2) \eta_{\bar{s}}(k_3) \psi_{\bar{s}\nu}(k_4),
\end{eqnarray}
where $\chi_A=1$ and $\chi_B=-1$ and the coupling constant $\Gamma$ is given by
\begin{equation}
\Gamma = J_K^2 \int_q \frac{t\Delta |\lambda(\bq)|^2}{(q_0^2+t^2|\lambda(\bq)|^2)(q_0^2+\Delta^2)}.
\end{equation}

While the ration $|J|/U$ is fixed, regardless of the strength of the Kondo coupling, the ratio $\Gamma/U$ depends on the ratio $c/v = |K|/t$ of Fermi velocities of  Majorana fermions 
and conduction electrons. The evolution of the interaction strength $U$, $|J|$ and $\Gamma$ in units of $J_K^2/|K|$ as a function of $c/v$ is shown in Fig.~\ref{figure2}.

In general, the Kitaev exchange between local moment spins is mediated through a super-exchange mechanism and therefore expected to be weak compared to typical 
electronic energy scales. For example, ferromagnetic Kitaev exchanges in the range 5-20 meV have been estimated for $\alpha$-RuCl$_3$ \cite{Yadav+16,Suzuk+21} 
and Na$_2$IrO$_3$ \cite{Katukuri+14}, while certain Co-based honeycomb materials were found to exhibit an antiferromagnetic Kitaev exchange of 2-3 meV \cite{Songvilay+20}.
For comparison, the nearest-neighbor tight-binding parameter for graphene is $t\approx  2.7$ eV \cite{Neto+09}. In real materials one should therefore expect a Fermi-velocity 
ratio $c/v$ of the order of $10^{-3}-10^{-2}$, implying that $\Gamma\ll U$. However, we will theoretically explore the full range of $c/v$ and demonstrate that even in the regime of 
very strong coupling $\Gamma$ between Majorana fermions and conduction electrons the critical behaviour will remain the same.

In addition to the interactions $U$, $J$, and $\Gamma$  we will also include a density-density interaction $\rho$ between 
conduction electrons on neighboring sites,
\begin{eqnarray}
S_\rho [\bar{\bm{\psi}},\bm{\psi}] & = &  \rho\sum_{s} \int_{k_1,\ldots,k_4} \delta_{k_1-k_2+k_3-k_4}\nn\\
& & \times \left( \bar{\bm{\psi}}_{s}(k_1)\bm{\psi}_{s}(k_2)\right) \left( \bar{\bm{\psi}}_{\bar{s}}(k_3) \bm{\psi}_{\bar{s}}(k_4)\right).
\end{eqnarray}
Although this interaction is initially zero, it will be generated by the other interactions under the RG.

\section{Renormalization-Group Analysis}
\label{sec.RG}

The quantum criticality of interacting Dirac semimetals is usually analyzed using a Gross-Neveu-Yukawa field theory
that describes the coupling of a dynamical order parameter field to gapless Dirac fermions \cite{Herbut06,Herbut+09}. 
However, in a metallic system with competing interactions it is often unclear what the leading ordering 
instability is. We therefore use a perturbative parquet RG analysis \cite{Murray+14,Chubukov+16,Xing+17,Parthenios+23}
to understand the scale dependence of the interaction parameters $U$, $J$, $\rho$ and $\Gamma$. Under the RG interaction 
parameters can depart significantly from their bare initial values. This can lead to attraction in unconventional superconducting 
pairing channels, as discussed in the context of iron-based superconductors \cite{Chubukov+16,Xing+17}. 

The four interactions $U$, $J$, $\rho$ and $\Gamma$ form a basis set for the subspace of interactions explored 
under the RG since it is not possible to express one of the interactions as a linear combination of the other three. However, by using 
a Fierz identity one may express the density-density interaction $\rho$ in terms of the spin-spin interaction $J$ and an additional 
pair-hopping vertex. Such ambiguities could lead to biased results when carrying out a Hubbard-Stratonovich or mean-field decoupling 
but wouldn’t affect our perturbative RG analysis.

\begin{figure}[t!]
 \includegraphics[width=\linewidth]{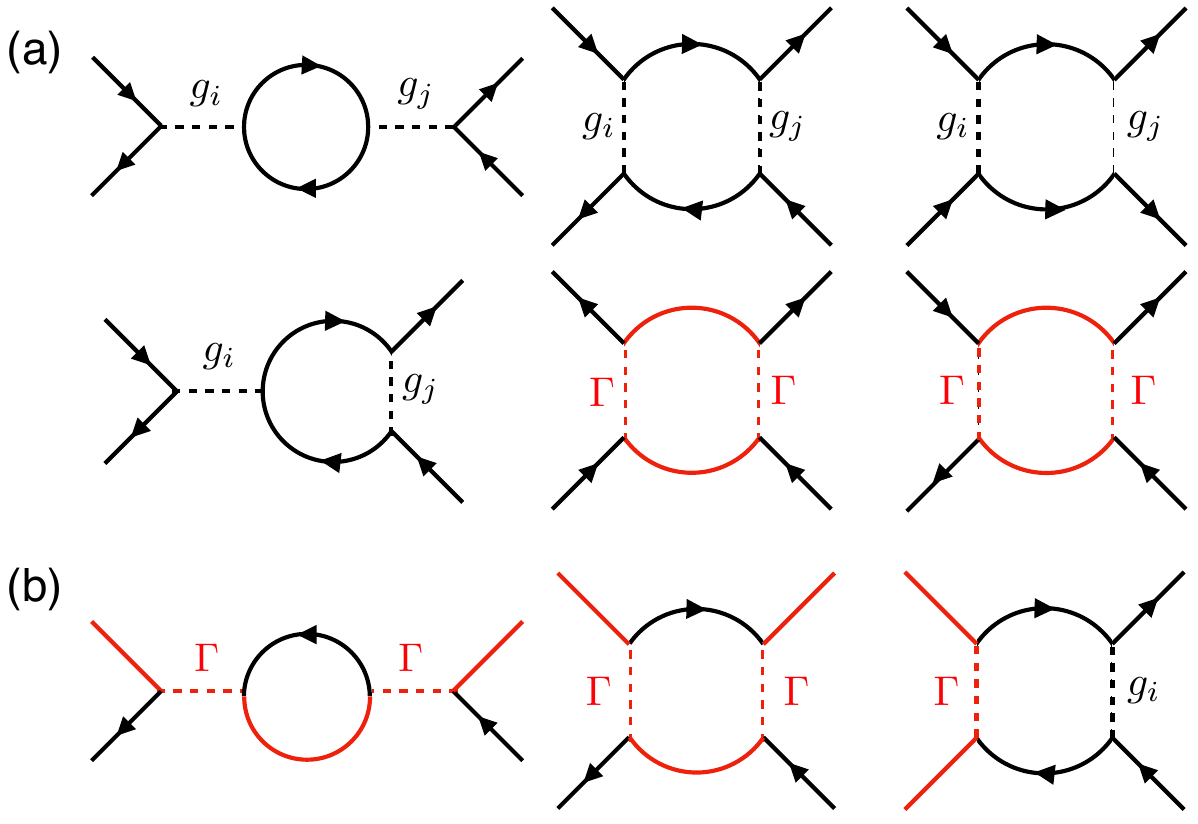}
 \caption{Second-order, one-loop diagrams that renormalize (a) the electron-electron interactions $g_i\in \{U, J, \rho \}$ and (b) the 
 coupling vertex $\Gamma$ between conduction electrons and Majorana fermions. Conduction electron fields $\bar{\bm{\psi}}$, $\bm{\psi}$
 correspond to black, Majorana fermion fields $\bm{\eta}$ to red lines.}
\label{figure3}
\end{figure}

To focus on the long-wavelength behavior and impose a momentum cut-off $|\bk|\le\Lambda$, where $\bk$ 
measure the distance for the Dirac points $\bK_\pm$. Under the perturbative RG scheme we integrate out modes with 
momenta from the infinitesimal shell $\Lambda e^{-d\ell} \le |\bk| \le \Lambda$, followed by a rescaling of frequency, 
$k_0'=k_0 e^{z d\ell}$, momenta, $\bk' = \bk e^{d\ell}$, and fields, $\bm{\psi}'(k')=\bm{\psi}(k)e^{(\Delta_\psi/2) d\ell}$, 
$\bm{\eta}'(k')=\bm{\eta}(k)e^{(\Delta_\eta/2) d\ell}$. Since the  inverse fermion propagators are not renormalized 
by contractions of the interaction vertices we can keep them scale invariant by setting $z=1$ and $\Delta_\psi =\Delta_\eta=-(d+2)$,
where $d$ is the spatial dimension. The resulting scaling dimensions of the interactions are equal to 
$[U]=[V]=[\rho]=[\Gamma] = -d+1$.

\begin{figure*}[t!]
 \includegraphics[width=\linewidth]{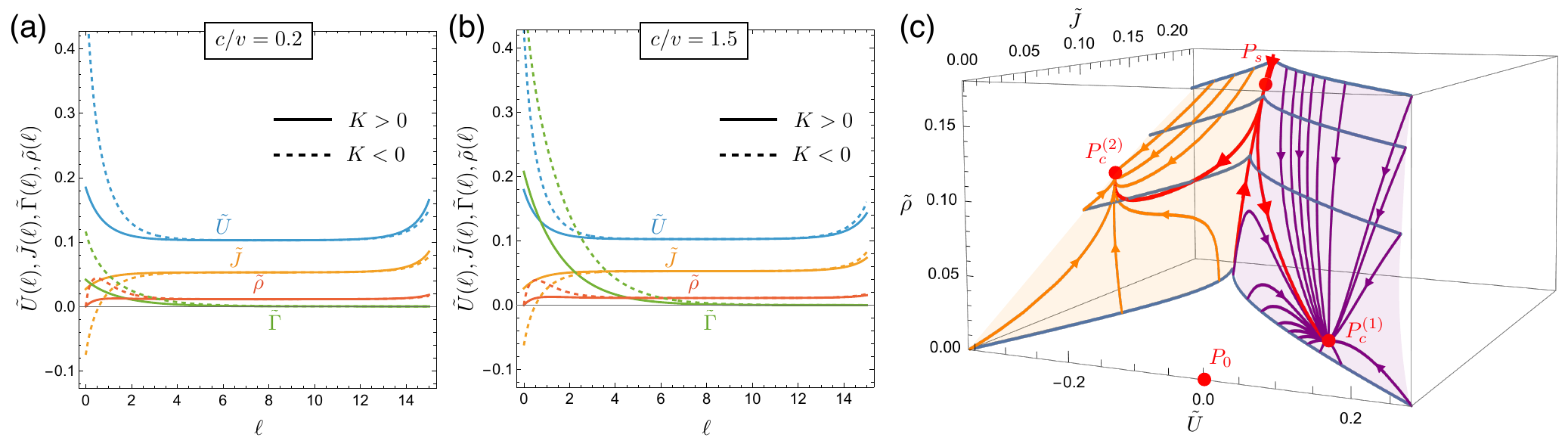}
 \caption{(a) RG flow of the rescaled interactions $\tilde{U}$, $\tilde{J}$, $\tilde{\rho}$, and $\tilde{\Gamma}$. The initial 
 interaction strengths are derived from the microscopic Kitaev-Kondo model with Fermi-velocity ratio $c/v=0.2$ and for both FM (dashed) and 
 AFM (solid) Kitaev couplings. The Kondo coupling is tuned very slightly above the critical value in both cases.  (b) Same as in (a) but for $c/v=1.5$. In all 
 cases the critical behavior is controlled by the same critical fixed point with $\tilde{\Gamma}_c=0$ and $\tilde{U}_c, \tilde{J}_c, \tilde{\rho}_c>0$, corresponding 
 to the plateau values. (c) Fixed points and critical surface for $\tilde{\Gamma}=0$. The trajectories show the RG flow within the critical surface. 
 The symmetry-breaking phase transition of the Kitaev-Kondo model is 
 controlled by the critical fixed point $P_c^{(1)}$.}
\label{figure4}
\end{figure*}

The above scaling analysis shows that in $d=2$ the interactions are irrelevant perturbations at the non-interacting FL$^*$ fixed 
points. This is expected because of the vanishing density of states at the Dirac points. Strictly speaking, the perturbative RG 
is controlled in $d=1+\epsilon$ where the interaction strengths at the critical fixed points will be of order $\epsilon$. 

 The one-loop diagrams that renormalize the electron-electron interactions $g_i \in \{U,J,\rho\}$ and the coupling vertex 
 $\Gamma$ between conduction electrons and gapless Majorana fermions are shown in Fig.~\ref{figure3} (a) and (b), respectively.  
 The required shell integrals can be easily calculated in $d=2$,
 \begin{eqnarray}
\int_q^> G^\psi_{s\bar{s}}(q) G^\psi_{\bar{s} s}(q) & = & \frac{1}{8\pi}\frac{\Lambda}{v} d\ell  \\
\int_q^> G^\psi_{ss}(q) G^\psi_{s's'}(q) & = & -\frac{1}{8\pi}\frac{\Lambda}{v} d\ell \\
\int_q^> G^\eta_{s\bar{s}}(q) G^\psi_{\bar{s}s}(q) & = & i\chi_s \frac{1}{8\pi}\frac{2\Lambda}{c+v} d\ell \\
\int_q^> G^\eta_{ss}(q) G^\eta_{s's'}(q) & = & -\frac{1}{8\pi}\frac{\Lambda}{c} d\ell
\end{eqnarray}
where $v$ and $c$ denote the velocities of Dirac electrons and Majorana fermions, respectively, and we have defined  
\begin{equation}
\int_q^>  = \frac{1}{(2\pi)^3} \int_{-\infty}^\infty dq_0 \int_{\Lambda e^{-d\ell} \le |\bq| \le \Lambda} d^2\bq,
\end{equation}
for brevity. The resulting RG equations for the rescaled interactions $\tilde{g}_i = \frac{1}{8\pi}\frac{\Lambda}{v} g_i$ and 
$\tilde{\Gamma} = \frac{1}{8\pi}\frac{\Lambda}{v} \Gamma$ read
\begin{eqnarray}
\label{eq.RG1}
\frac{d\tilde{U}}{d\ell} & = & -\tilde{U} -4\tilde{\rho}^2+12\tilde{J}^2+4\tilde{U}(\tilde{\rho}+3\tilde{J}),\\
\label{eq.RG2}
\frac{d\tilde{J}}{d\ell} & = & -\tilde{J} +\tilde{U}^2 -\tilde{\rho}^2+7\tilde{J}^2+2\tilde{J}(2\tilde{U}+\tilde{\rho})+\frac{v}{4c}\tilde{\Gamma}^2,\\
\label{eq.RG3}
\frac{d\tilde{\rho}}{d\ell} & = & -\tilde{\rho} +\tilde{U}^2 +3\tilde{\rho}^2+3\tilde{J}^2-2\tilde{\rho}(3\tilde{J}+2\tilde{U})+\frac{v}{4c}\tilde{\Gamma}^2,\\
\label{eq.RG4}
\frac{d\tilde{\Gamma}}{d\ell} & = & -\tilde{\Gamma} + \frac{2v}{c+v}\tilde{\Gamma}^2 +2\tilde{\Gamma}(\tilde{\rho}+3\tilde{J}).
\end{eqnarray}

As a first step we numerically integrate the RG equations, starting with different initial values $\tilde{g}_i (0)$ and $\tilde{\Gamma}(0)$. As discussed in Sec.~\ref{sec.low_en}, the 
ratio $|\tilde{J}(0)|/\tilde{U}(0) \approx 0.146$ is fixed while the ratio $\tilde{\Gamma}(0)/\tilde{U}(0)$ is a function of the velocity ratio $c/v$ (see Fig.~\ref{figure2}).  
The bare value of the density-density interaction is zero, $\tilde{\rho}(0)=0$. The sign of $\tilde{J}(0)$ can be positive or negative, depending on the 
sign of the microscopic Kitaev exchange. The overall energy scale of the bare interactions can be tuned by the Kondo coupling $J_K$. For small $J_K$ all interactions 
will renormalize to zero, corresponding to the FL$^*$ phase. For $J_K$ above a critical value at least  some interactions will diverge, indicative of symmetry breaking.  

In Fig.~\ref{figure4} (a) and (b) we show the scale dependent interactions for different values of $c/v$ and both, 
positive and negative signs of $\tilde{J}(0)$. In all cases we use bisection to find $J_K$ slightly above and infinitesimally close to the critical value. This means 
that over a large range of scales $\ell$ the trajectories will be stalled very close to a critical fixed point, corresponding to plateaus of  $\tilde{g}_i (\ell)$, $\tilde{\Gamma}(\ell)$.

We find that $\tilde{\Gamma}(\ell)$ always renormalizes to zero, even for large $c/v$ where $\tilde{\Gamma}$ is initially the largest interaction. This means that the 
Majorana fermions  decouple from the conduction electrons. Moreover, a positive interaction $\tilde{\rho}$ is generated under the RG. Interestingly, the critical behavior is 
always controlled by the same fixed point, irrespective of the ratio $c/v$ and the initial sign of $\tilde{J}$. At the critical fixed point we obtain $\tilde{\Gamma}=0$, 
$\tilde{U}_c\approx 0.1027$, $\tilde{J}_c\approx 0.053$, and $\tilde{\rho}_c\approx 0.011$ from the plateau values. 

To better understand the critical behavior we analyze the RG equations for $\tilde{\Gamma}=0$. In addition to the trivial 
non-interacting FL$^*$ fixed point $P_0$, the coupled RG equations for $\tilde{U}$, $\tilde{J}$, $\tilde{\rho}$  exhibit three non-trivial fixed 
points $P_c^{(1)}$, $P_c^{(2)}$, and $P_s$, which are all in the domain $\tilde{\rho}>0$ and $\tilde{J}\ge0$ (see Fig.~\ref{figure4}(c)). The non-trivial fixed 
points are located on the critical surface that separates the FL$^*$ phase at weak interactions from the phases where interactions diverge under the perturbative RG 
scheme, signaling some form of symmetry breaking. $P_c^{(1)}$ and  $P_c^{(2)}$ are located on different sheets of the critical surface, shaded in orange and purple in Fig.~\ref{figure4}(c), 
and are stable fixed points along the tangential directions. They are the 
critical fixed points that correspond to different types of symmetry breaking. 
The metastable fixed point $P_s$ is located on the separatrix between the two sheets. 

From the plateau values of $\tilde{g}_i (\ell)$ (see Fig.~\ref{figure4} (a), (b)) it is clear that the critical fixed point that controls the symmetry breaking transition of the Kitaev-Kondo 
model is given by $P_c^{(1)}$ at  
\begin{equation}
\tilde{U}_c  =  \frac{2\sqrt{3}-1}{24},\;\; \tilde{J}_c  =  \frac{3- \sqrt{3}}{24}\;\;\textrm{and}\;\; \tilde{\rho}_c =   \frac{2- \sqrt{3}}{24}.
\label{eq.crit}
\end{equation}

\section{Field renormalization and susceptibility exponents}
\label{sec.fieldRG}

Since all three electron-electron interactions are finite at the critical fixed point $P_c^{(1)}$ it requires a more careful analysis to identify the type of 
symmetry breaking that occurs at the phase transition. While sufficiently strong positive $\tilde{\rho}$ and $\tilde{U}$ favor CDW and SDW orders, respectively, 
it was found that a positive $\tilde{J}$ can lead to an instability towards $d+id$ superconductivity, at least away from half-filling \cite{Bunney+25}.

The susceptibility exponent $\gamma$ for a given order parameter can be obtained from the field scaling exponent $y_h$ of the conjugate field $h$, using the 
relation $\gamma = (2 y_h-D)\nu$, where $D$ is the dimension and $\nu$ the correlation-length exponent \cite{Cardy_1996}. For the quantum critical point of the 
Kitaev-Kondo model, $D=2+1$. 

\begin{figure}[t!]
 \includegraphics[width=\linewidth]{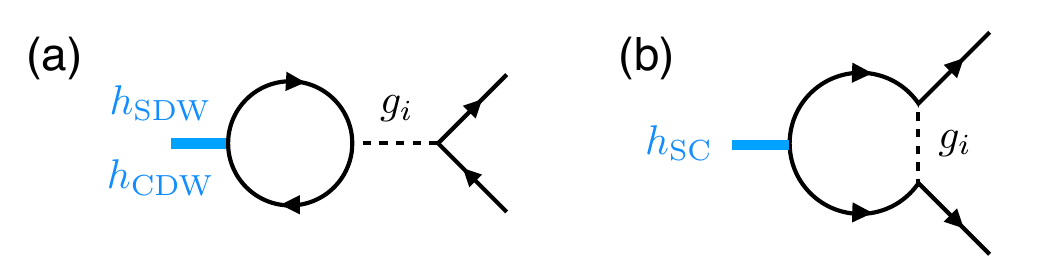}
 \caption{Diagrams that contribute to the renormalization of the fields (a) $h_\textrm{CDW}$, $h_\textrm{SDW}$ and (b) $h_\textrm{SC}$ to linear order.}
\label{figure5}
\end{figure}

The correlation length exponent $\nu$ can be obtained from the divergence of 
the interaction parameters. Consider for example a simple situation where all interactions  diverge in the same way, corresponding to linearized RG 
equations $\delta_i'(\ell) = b \delta_i(\ell)$ for $\delta_i = \tilde{g}_i-\tilde{g}_{i,c}$ with some coefficient $b>0$, resulting in $\delta_i(\ell) = \delta_i(0) e^{b\ell}$.
We can define the correlation length $\xi \sim e^{\ell^*}$ from the scale $\ell^*$ where the interactions become of order one, $\delta_i(\ell^*)\simeq 1$. 
This results in $\xi \sim \delta_i(0)^{-1/b}$ and hence $\nu=1/b$.

In the present situation, linearizing the RG equations (\ref{eq.RG1})-(\ref{eq.RG3}) for $\tilde{g}_i \in \{ \tilde{U}, \tilde{J}, \tilde{\rho} \}$ near $P_c^{(1)}$ leads to coupled 
linear equations, $\delta_i'(\ell) = \sum_j = A_{ij} \delta_j(\ell)$, where the matrix $\bA$ is completely determined by the critical values (\ref{eq.crit}). The correlation 
length $\nu$ is given by the inverse of the largest eigenvalue of $\bA$, resulting in $\nu=1$. 

We first investigate how the fields conjugate to CDW and SDW order renormalize. The corresponding field terms in the continuum field theory are given by
\begin{eqnarray}
\mathcal{S}_{h,\textrm{CDW}}[\bar{\bm{\psi}},\bm{\psi}] & = &  -h_{\textrm{CDW}} \int_k \bar{\bm{\psi}}_k\left( \bm{\tau}_z \otimes \bm{1} \right)\bm{\psi}_k,\\
\mathcal{S}_{h,\textrm{SDW}}[\bar{\bm{\psi}},\bm{\psi}] & = & -h_{\textrm{SDW}} \int_k \bar{\bm{\psi}}_k \left( \bm{\tau}_z \otimes \bm{\sigma}_z \right) \bm{\psi}_k,
\end{eqnarray}
where $\bm{\tau}_z$ and $\bm{\sigma}_z$ denote Pauli matrices in sub-lattice and spin space, respectively. 
The diagrams that contribute to the field renormalization to linear order are shown in Fig.~\ref{figure5}(a), resulting in
\begin{eqnarray}
\label{eq.field1}
\frac{d h_{\textrm{CDW}}}{d\ell} & = & (1-4\tilde{U} +8 \tilde{\rho} ) h_{\textrm{CDW}}, \\
\label{eq.field2}
\frac{d h_{\textrm{SDW}}}{d\ell} & = & (1+4\tilde{U} +8 \tilde{J} ) h_{\textrm{SDW}}.
\end{eqnarray}

It was found that for an antiferromagnetic $\tilde{J}$ the leading SC instability of the lattice model away from half filling is in the spin-singlet channel across neighboring sites 
with a spatial  $d+id$ structure, described by the pairing term \cite{Bunney+25}
\begin{eqnarray}
\hat{\mathcal{H}}_{\textrm{SC}} & = &  \sum_{\gamma=x,y,z} \sum_{\langle i,j\rangle_\gamma} \Delta_\gamma \sum_{\nu,\nu'} \hat{c}^\dagger_{i\nu} (i\bm{\sigma}_y)_{\nu\nu'}\hat{c}^\dagger_{j\nu'}
+\textrm{h.c.}\\
& = & \sum_{\bk,\gamma}\Delta_\gamma e^{i\bk\bm{\delta}_\gamma} \sum_{\nu,\nu'} \hat{c}^\dagger_{A\nu}(\bk) (i\bm{\sigma}_y)_{\nu\nu'}\hat{c}^\dagger_{B\nu'}(-\bk)+\textrm{h.c.},\nn
\end{eqnarray}
with SC order parameter $\vec{\Delta}  =  \Delta \vec{d}$, 
\begin{equation}
\vec{d}=\vec{d}_{x^2-y^2}+i \vec{d}_{xy} = \frac{1}{\sqrt{6}} \left( \begin{array}{c} -1 \\ -1 \\ 2   \end{array}  \right)  
+\frac{i}{\sqrt{2}} \left( \begin{array}{c} 1 \\ -1 \\ 0   \end{array}  \right).
\end{equation}

To obtain the effective pairing term in the low-energy continuum field theory, we expand around the Dirac point $\bK_+$ and keep only 
the leading term, $\sum_\gamma \Delta_\gamma e^{i\bk\bm{\delta}_\gamma}\approx \sqrt{6} \Delta$. The resulting SC field term is given by
\begin{eqnarray}
\mathcal{S}_{h,\textrm{SC}}[\bar{\bm{\psi}},\bm{\psi}] & = & -h_{\textrm{SC}} \int_k \left\{\bar{\bm{\psi}}_k \left[\bm{\tau}_x \otimes (i \bm{\sigma}_y)\right]  \bar{\bm{\psi}}_{-k}^T \right.\nn\\
& & + \left. \bm{\psi}^T_k \left[\bm{\tau}_x \otimes (-i \bm{\sigma}_y)\right]  \bm{\psi}_{-k}  \right\},
\end{eqnarray}
and the renormalization of $h_\textrm{SC}$ to linear order obtained from the diagrams in Fig.~\ref{figure5}(b), resulting in 
\begin{equation}
\label{eq.field3}
\frac{d h_{\textrm{SC}}}{d\ell}  =  (1+6\tilde{J} -2 \tilde{\rho} ) h_{\textrm{SC}}. \\
\end{equation}

Evaluating the field RG equations (\ref{eq.field1}), (\ref{eq.field2}), and (\ref{eq.field3}) at the critical point $P_c^{(1)}$ we obtain the field scaling exponents $y_h$ and from 
 the relation $\gamma = 2y_h-3$  the susceptibility exponents
\begin{eqnarray}
\gamma_{\textrm{CDW}} & = & -1-8\tilde{U}_c+16\tilde{\rho}_c = -\frac23 (2\sqrt{3}-1),\\
\gamma_{\textrm{SDW}} & = & -1+8\tilde{U}_c+16\tilde{J}_c = \frac23, \\
\gamma_{\textrm{SC}} & = & -1+12\tilde{J}_c-4\tilde{\rho}_c = -\frac16 (2\sqrt{3}-1).
\end{eqnarray}
Since only $\gamma_{\textrm{SDW}}$ is positive, we conclude that the FL$^*$ phase at half filling becomes unstable towards the formation of SDW order.

 The Landau free-energy analysis of different SC order parameters of Ref.~\cite{Bunney+25} is fairly general and should also apply to electron fillings close to the 
Dirac point. One would therefore expect that the chiral $d\pm id$ state remains the supeconducting state with the lowest free energy at half filling. 
However, for completeness, we have checked superconducting states with other pairing symmetries. As one might expect, we found that triplet pairing is disfavoured by 
antiferromagnetic $J$, while  on-site singlet pairing is strongly suppressed by the Hubbard repulsion $U$.

\section{Discussion}
\label{sec.disc}

We have analyzed the instabilities of the fractionalized Fermi liquid that arrises in a Kondo lattice model of Dirac conduction electrons on the half-filled 
honeycomb lattice coupled to a Kitaev QSL. After fractionalizing the spin-1/2 operators into Majorana fermions, we have used second-order perturbation 
theory to obtain the interactions between the gapless fermion modes in the effective low-energy, continuum field theory. 
 The bare theory contains three different interaction vertices, a Hubbard repulsion $U$ and a spin-spin interaction $J$ between 
Dirac electrons, as well as a coupling $\Gamma$ between conduction electrons and Dirac Majorana fermions. We have analyzed the scale dependence of the interactions, using perturbative 
parquet RG. In addition to the aforementioned interactions, an electronic density-density repulsion $\rho$ is generated. 

Regardless of the ratio of Fermi velocities and the sign of the Kitaev coupling, the RG flow is always controlled by the same critical fixed point, at which the 
Dirac electrons decouple from the Majorana fermions and all three electron-electron interaction have finite positive values. 
This results suggests that it is justified to  integrate out the Majorana fermions entirely, as done in Ref.~\cite{Bunney+25}. 
However, we stress that the velocity ratio $c/v$ could potentially renormalize beyond one-loop order, resulting in a different flow of 
the coupling $\Gamma$ between conduction electrons and Dirac Majorana fermions. Velocity renormalization is a well known phenomenon of strongly coupled GNY 
theories and perturbative RG, even beyond one-loop order,  cannot unambiguously rule out the existence of a novel GNY critical fixed point with finite coupling to Majorana fermions.

To determine the type of symmetry breaking  
we have derived the RG equations of the fields conjugate to CDW, SDW and SC order, from which we determined the field scaling exponents at 
the critical point and the susceptibility exponents. For the SC order parameter we have focused on singlet pairing across neighboring sites with a spatial 
$d+id$ form factor \cite{Bunney+25}. While the antiferromagnetic spin-spin interaction $J$ stabilizes antiferromagnetism and SC in almost equal terms, the AFM ordering is strongly enhanced
by the Hubbard repulsion $U$ whereas SC is destabilized by the repulsive density-density interaction $\rho$. Our results show that the only divergent susceptibility is toward antiferromagnetic (SDW) order. 

While the perturbative parquet RG is the method of choice to identify the leading symmetry-breaking instability in the presence of multiple competing 
interactions \cite{Murray+14,Chubukov+16,Xing+17,Parthenios+23}, the critical exponents for the correlation length $\nu=1$ and SDW susceptibility $\gamma=2/3$ 
obtained this way are only crude estimates. Ultimately, the SDW transition at half filling should falls into the Gross-Neveu-Heisenberg universality class in 2+1 dimensions. 
From interpolation between series expansions near lower and upper critical dimensions \cite{Ladovrechis+23} with previous results from 
fourth-order $4-\epsilon$ \cite{Zerf+17} and second-order $1/N$ \cite{Gracey18} expansions, functional RG \cite{Janssen+14,Knorr18},  as well as 
quantum Monte Carlo simulations \cite{Parisen+15,Liu+19,Liu+21,Otsuka+16,Otsuka+20,Xu+21,Buividovich+18,Buividovich+19,Ostmeyer+20,Ostmeyer+21} 
estimates of $1/\nu \approx 0.83$ for the inverse correlation length exponent and $\eta_\phi\approx 1.01$ for the anomalous dimension of the order parameter field 
were obtained for $N=4$ (valley and spin) 2-component Dirac fermion fields \cite{Ladovrechis+23}. 
Assuming hyperscaling, this would imply $\gamma = (2-\eta_\phi)\nu \approx 1.19$. 

Our analysis is based on an effective continuum field theory and neglects lattice effects. According to the Landau cubic criterion, three-fold clock terms, relevant for the honeycomb lattice, would render the transition first order. While such lattice terms are indeed relevant at the Wilson-Fisher fixed point, they turn irrelevant at fermion-induced critical points 
described by GNY field theories \cite{Li+17,Christou+20}. This does not rule out however, that fluctuations from shorter length scales induce first-order behavior before the coarse grained continuum 
description becomes applicable. 

Our results are not in direct contradiction with previous studies of the Kitaev-Kondo model on the honeycomb lattice. Mean field investigations \cite{Seifert+18,Choi+18} found 
that for a FM Kitaev coupling and electron fillings away from the Dirac point a $p$-wave SC state forms between the FL$^*$ phase and the heavy FL. On approaching half filling, the 
FL$^*$/SC transition moves towards stronger coupling and the region of superconductivity shrinks to almost zero \cite{Seifert+18}. At half filling one would expect that fluctuations 
beyond mean-field theory play a crucial role. The strong renormalization of the interactions parameters indeed confirms this. Most notable are the fluctuation-driven generation of the 
density-density repulsion $\rho$, which disfavors SC, and the sign change of the spin interaction $J$ from FM to AFM.

In a recent study \cite{Bunney+25} of the AFM Kitaev-Kondo model all degrees of freedom associated with the Kitaev sector were integrated out and 
the resulting electronic Hamiltonian analyzed using functional RG (fRG).  The fRG found an instability 
toward a SDW at electron fillings close to the van-Hove singularity and a transition to a $d+id$ SC state at lower fillings, but still far above the Dirac points. 
Unfortunately, it was not possible to analyze the behavior at fillings close to the Dirac point, due to finite momentum resolution \cite{Bunney+25}. It is likely that the AFM state only 
exists very close to half-filling and that lattice effects, which are included in the fRG, play an essential role in stabilizing exotic superconductivity at larger fillings. On the lattice, the 
electronic spin interactions inherit the bond-directional dependence of the Kitaev exchange. This frustration is likely to destabilize AFM order.  

 To better understand the instabilities of different types of fractionalized Fermi liquids towards exotic superconductivity and other broken symmetry states 
it will be important to investigate other toy models that host  fractionalized FL$^*$ phases. 
In a very recent study \cite{Tang+25} it was demonstrated that when the so called “Yao-Lee Model”  \cite{Yao+11}, a spin-orbital generalisation of the Kitaev model, is 
Kondo-coupled to conduction electrons, the exchange of Majorana spinons can drive a (fractionalized) topological $d+id$ pairing instability in the weak-coupling regime. 
In future studies it would be interesting to allow for Kondo coupling not only to the SU(2) symmetric local moment spins but also to the Kitaev orbital pseudo-spin sector and to systematically investigate the dependence on the electron filling.

A frustrated Kondo lattice model with an FL$^*$ phase that is s amenable to sign free auxiliary-field quantum Monte Carlo simulations 
was constructed \cite{Hofmann+19} by coupling Dirac electrons on the honeycomb lattice to spin-1/2 degrees of freedom 
on the Kagome lattice. The interactions between the spins are chosen along the lines of the Balents-Fisher-Girvin model \cite{Balents+02} which is known to host a 
$Z_2$ spin-liquid and a ferromagnetic phase. While the authors characterized the FL$^*$ phase in great detail through spectral functions and mutual information 
between electrons and spins, they did not investigate possible superconducting instabilities.

Studies of various toy models have shown that fractionalized Fermi liquids naturally arise  when quantum spin liquids are weakly coupled to conduction electrons in 
a Kondo lattice setting. The emergent fermions in such FL$^*$ phases mediate effective electron-electron interactions that can drive instabilities towards exotic 
superconductivity or other symmetry-broken phases. The emergent fermions therefore act as pairing glue similar to the phonons in conventional BCS superconductivity. Our
work shows that at the quantum critical point of the FL$^*$ phase of the half-filled Kitaev-Kondo model the emergent Majorana fermions decouple from the conduction 
electrons. It remains an intriguing question wether this behavior is generic or if certain models can support quantum critical points at which emergent fermions and conduction 
electrons remain strongly coupled, opening up the possibility for the formation of even more exotic quantum phases.

\acknowledgements

J.L. is grateful for discussions with Cristian Batista, Claudio Castelnovo, and Cecilie Glittum. F.K. acknowledges fruitful discussions with Chris Hooley,
Lukas Janssen, and Hong Yao.


\begin{thebibliography}{47}%
\makeatletter
\providecommand \@ifxundefined [1]{%
 \@ifx{#1\undefined}
}%
\providecommand \@ifnum [1]{%
 \ifnum #1\expandafter \@firstoftwo
 \else \expandafter \@secondoftwo
 \fi
}%
\providecommand \@ifx [1]{%
 \ifx #1\expandafter \@firstoftwo
 \else \expandafter \@secondoftwo
 \fi
}%
\providecommand \natexlab [1]{#1}%
\providecommand \enquote  [1]{``#1''}%
\providecommand \bibnamefont  [1]{#1}%
\providecommand \bibfnamefont [1]{#1}%
\providecommand \citenamefont [1]{#1}%
\providecommand \href@noop [0]{\@secondoftwo}%
\providecommand \href [0]{\begingroup \@sanitize@url \@href}%
\providecommand \@href[1]{\@@startlink{#1}\@@href}%
\providecommand \@@href[1]{\endgroup#1\@@endlink}%
\providecommand \@sanitize@url [0]{\catcode `\\12\catcode `\$12\catcode
  `\&12\catcode `\#12\catcode `\^12\catcode `\_12\catcode `\%12\relax}%
\providecommand \@@startlink[1]{}%
\providecommand \@@endlink[0]{}%
\providecommand \url  [0]{\begingroup\@sanitize@url \@url }%
\providecommand \@url [1]{\endgroup\@href {#1}{\urlprefix }}%
\providecommand \urlprefix  [0]{URL }%
\providecommand \Eprint [0]{\href }%
\providecommand \doibase [0]{https://doi.org/}%
\providecommand \selectlanguage [0]{\@gobble}%
\providecommand \bibinfo  [0]{\@secondoftwo}%
\providecommand \bibfield  [0]{\@secondoftwo}%
\providecommand \translation [1]{[#1]}%
\providecommand \BibitemOpen [0]{}%
\providecommand \bibitemStop [0]{}%
\providecommand \bibitemNoStop [0]{.\EOS\space}%
\providecommand \EOS [0]{\spacefactor3000\relax}%
\providecommand \BibitemShut  [1]{\csname bibitem#1\endcsname}%
\let\auto@bib@innerbib\@empty
\bibitem [{\citenamefont {Anderson}(1973)}]{Anderson73}%
  \BibitemOpen
  \bibfield  {author} {\bibinfo {author} {\bibfnamefont {P.}~\bibnamefont
  {Anderson}},\ }\bibfield  {title} {\bibinfo {title} {{Resonating valence
  bonds: A new kind of insulator?}},\ }\href
  {https://doi.org/https://doi.org/10.1016/0025-5408(73)90167-0} {\bibfield
  {journal} {\bibinfo  {journal} {Materials Research Bulletin}\ }\textbf
  {\bibinfo {volume} {8}},\ \bibinfo {pages} {153} (\bibinfo {year}
  {1973})}\BibitemShut {NoStop}%
\bibitem [{\citenamefont {Balents}(2010)}]{Balents10}%
  \BibitemOpen
  \bibfield  {author} {\bibinfo {author} {\bibfnamefont {L.}~\bibnamefont
  {Balents}},\ }\bibfield  {title} {\bibinfo {title} {{Spin liquids in
  frustrated magnets}},\ }\href {https://doi.org/10.1038/nature08917}
  {\bibfield  {journal} {\bibinfo  {journal} {Nature}\ }\textbf {\bibinfo
  {volume} {464}},\ \bibinfo {pages} {199} (\bibinfo {year}
  {2010})}\BibitemShut {NoStop}%
\bibitem [{\citenamefont {Senthil}\ \emph {et~al.}(2003)\citenamefont
  {Senthil}, \citenamefont {Sachdev},\ and\ \citenamefont
  {Vojta}}]{Senthil+03}%
  \BibitemOpen
  \bibfield  {author} {\bibinfo {author} {\bibfnamefont {T.}~\bibnamefont
  {Senthil}}, \bibinfo {author} {\bibfnamefont {S.}~\bibnamefont {Sachdev}},\
  and\ \bibinfo {author} {\bibfnamefont {M.}~\bibnamefont {Vojta}},\ }\bibfield
   {title} {\bibinfo {title} {{Fractionalized Fermi Liquids}},\ }\href
  {https://doi.org/10.1103/PhysRevLett.90.216403} {\bibfield  {journal}
  {\bibinfo  {journal} {Phys. Rev. Lett.}\ }\textbf {\bibinfo {volume} {90}},\
  \bibinfo {pages} {216403} (\bibinfo {year} {2003})}\BibitemShut {NoStop}%
\bibitem [{\citenamefont {Senthil}\ \emph {et~al.}(2004)\citenamefont
  {Senthil}, \citenamefont {Vojta},\ and\ \citenamefont
  {Sachdev}}]{Senthil+04}%
  \BibitemOpen
  \bibfield  {author} {\bibinfo {author} {\bibfnamefont {T.}~\bibnamefont
  {Senthil}}, \bibinfo {author} {\bibfnamefont {M.}~\bibnamefont {Vojta}},\
  and\ \bibinfo {author} {\bibfnamefont {S.}~\bibnamefont {Sachdev}},\
  }\bibfield  {title} {\bibinfo {title} {{Weak magnetism and non-Fermi liquids
  near heavy-fermion critical points}},\ }\href
  {https://doi.org/10.1103/PhysRevB.69.035111} {\bibfield  {journal} {\bibinfo
  {journal} {Phys. Rev. B}\ }\textbf {\bibinfo {volume} {69}},\ \bibinfo
  {pages} {035111} (\bibinfo {year} {2004})}\BibitemShut {NoStop}%
\bibitem [{\citenamefont {Hofmann}\ \emph {et~al.}(2019)\citenamefont
  {Hofmann}, \citenamefont {Assaad},\ and\ \citenamefont
  {Grover}}]{Hofmann+19}%
  \BibitemOpen
  \bibfield  {author} {\bibinfo {author} {\bibfnamefont {J.~S.}\ \bibnamefont
  {Hofmann}}, \bibinfo {author} {\bibfnamefont {F.~F.}\ \bibnamefont
  {Assaad}},\ and\ \bibinfo {author} {\bibfnamefont {T.}~\bibnamefont
  {Grover}},\ }\bibfield  {title} {\bibinfo {title} {{Fractionalized Fermi
  liquid in a frustrated Kondo lattice model}},\ }\href
  {https://doi.org/10.1103/PhysRevB.100.035118} {\bibfield  {journal} {\bibinfo
   {journal} {Phys. Rev. B}\ }\textbf {\bibinfo {volume} {100}},\ \bibinfo
  {pages} {035118} (\bibinfo {year} {2019})}\BibitemShut {NoStop}%
\bibitem [{\citenamefont {Qi}\ and\ \citenamefont {Sachdev}(2010)}]{Qi+10}%
  \BibitemOpen
  \bibfield  {author} {\bibinfo {author} {\bibfnamefont {Y.}~\bibnamefont
  {Qi}}\ and\ \bibinfo {author} {\bibfnamefont {S.}~\bibnamefont {Sachdev}},\
  }\bibfield  {title} {\bibinfo {title} {{Effective theory of Fermi pockets in
  fluctuating antiferromagnets}},\ }\href
  {https://doi.org/10.1103/PhysRevB.81.115129} {\bibfield  {journal} {\bibinfo
  {journal} {Phys. Rev. B}\ }\textbf {\bibinfo {volume} {81}},\ \bibinfo
  {pages} {115129} (\bibinfo {year} {2010})}\BibitemShut {NoStop}%
\bibitem [{\citenamefont {Moon}\ and\ \citenamefont {Sachdev}(2011)}]{Moon+11}%
  \BibitemOpen
  \bibfield  {author} {\bibinfo {author} {\bibfnamefont {E.~G.}\ \bibnamefont
  {Moon}}\ and\ \bibinfo {author} {\bibfnamefont {S.}~\bibnamefont {Sachdev}},\
  }\bibfield  {title} {\bibinfo {title} {{Underdoped cuprates as fractionalized
  Fermi liquids: Transition to superconductivity}},\ }\href
  {https://doi.org/10.1103/PhysRevB.83.224508} {\bibfield  {journal} {\bibinfo
  {journal} {Phys. Rev. B}\ }\textbf {\bibinfo {volume} {83}},\ \bibinfo
  {pages} {224508} (\bibinfo {year} {2011})}\BibitemShut {NoStop}%
\bibitem [{\citenamefont {Mei}\ \emph {et~al.}(2012)\citenamefont {Mei},
  \citenamefont {Kawasaki}, \citenamefont {Zheng}, \citenamefont {Weng},\ and\
  \citenamefont {Wen}}]{Mei+12}%
  \BibitemOpen
  \bibfield  {author} {\bibinfo {author} {\bibfnamefont {J.-W.}\ \bibnamefont
  {Mei}}, \bibinfo {author} {\bibfnamefont {S.}~\bibnamefont {Kawasaki}},
  \bibinfo {author} {\bibfnamefont {G.-Q.}\ \bibnamefont {Zheng}}, \bibinfo
  {author} {\bibfnamefont {Z.-Y.}\ \bibnamefont {Weng}},\ and\ \bibinfo
  {author} {\bibfnamefont {X.-G.}\ \bibnamefont {Wen}},\ }\bibfield  {title}
  {\bibinfo {title} {{Luttinger-volume violating Fermi liquid in the pseudogap
  phase of the cuprate superconductors}},\ }\href
  {https://doi.org/10.1103/PhysRevB.85.134519} {\bibfield  {journal} {\bibinfo
  {journal} {Phys. Rev. B}\ }\textbf {\bibinfo {volume} {85}},\ \bibinfo
  {pages} {134519} (\bibinfo {year} {2012})}\BibitemShut {NoStop}%
\bibitem [{\citenamefont {Anderson}(1987)}]{Anderson87}%
  \BibitemOpen
  \bibfield  {author} {\bibinfo {author} {\bibfnamefont {P.~W.}\ \bibnamefont
  {Anderson}},\ }\bibfield  {title} {\bibinfo {title} {{The Resonating Valence
  Bond State in La$_2$CuO$_4$ and Superconductivity}},\ }\href
  {https://doi.org/10.1126/science.235.4793.1196} {\bibfield  {journal}
  {\bibinfo  {journal} {Science}\ }\textbf {\bibinfo {volume} {235}},\ \bibinfo
  {pages} {1196} (\bibinfo {year} {1987})}\BibitemShut {NoStop}%
\bibitem [{\citenamefont {Seifert}\ \emph {et~al.}(2018)\citenamefont
  {Seifert}, \citenamefont {Meng},\ and\ \citenamefont {Vojta}}]{Seifert+18}%
  \BibitemOpen
  \bibfield  {author} {\bibinfo {author} {\bibfnamefont {U.~F.~P.}\
  \bibnamefont {Seifert}}, \bibinfo {author} {\bibfnamefont {T.}~\bibnamefont
  {Meng}},\ and\ \bibinfo {author} {\bibfnamefont {M.}~\bibnamefont {Vojta}},\
  }\bibfield  {title} {\bibinfo {title} {{Fractionalized Fermi liquids and
  exotic superconductivity in the Kitaev-Kondo lattice}},\ }\href
  {https://doi.org/10.1103/PhysRevB.97.085118} {\bibfield  {journal} {\bibinfo
  {journal} {Phys. Rev. B}\ }\textbf {\bibinfo {volume} {97}},\ \bibinfo
  {pages} {085118} (\bibinfo {year} {2018})}\BibitemShut {NoStop}%
\bibitem [{\citenamefont {Choi}\ \emph {et~al.}(2018)\citenamefont {Choi},
  \citenamefont {Klein}, \citenamefont {Rosch},\ and\ \citenamefont
  {Kim}}]{Choi+18}%
  \BibitemOpen
  \bibfield  {author} {\bibinfo {author} {\bibfnamefont {W.}~\bibnamefont
  {Choi}}, \bibinfo {author} {\bibfnamefont {P.~W.}\ \bibnamefont {Klein}},
  \bibinfo {author} {\bibfnamefont {A.}~\bibnamefont {Rosch}},\ and\ \bibinfo
  {author} {\bibfnamefont {Y.~B.}\ \bibnamefont {Kim}},\ }\bibfield  {title}
  {\bibinfo {title} {{Topological superconductivity in the Kondo-Kitaev
  model}},\ }\href {https://doi.org/10.1103/PhysRevB.98.155123} {\bibfield
  {journal} {\bibinfo  {journal} {Phys. Rev. B}\ }\textbf {\bibinfo {volume}
  {98}},\ \bibinfo {pages} {155123} (\bibinfo {year} {2018})}\BibitemShut
  {NoStop}%
\bibitem [{\citenamefont {Kitaev}(2006)}]{Kitaev06}%
  \BibitemOpen
  \bibfield  {author} {\bibinfo {author} {\bibfnamefont {A.}~\bibnamefont
  {Kitaev}},\ }\bibfield  {title} {\bibinfo {title} {{Anyons in an exactly
  solved model and beyond}},\ }\href
  {https://doi.org/10.1016/j.aop.2005.10.005} {\bibfield  {journal} {\bibinfo
  {journal} {Annals of Physics}\ }\textbf {\bibinfo {volume} {321}},\ \bibinfo
  {pages} {2–111} (\bibinfo {year} {2006})}\BibitemShut {NoStop}%
\bibitem [{\citenamefont {Bunney}\ \emph {et~al.}(2025)\citenamefont {Bunney},
  \citenamefont {Seifert}, \citenamefont {Rachel},\ and\ \citenamefont
  {Vojta}}]{Bunney+25}%
  \BibitemOpen
  \bibfield  {author} {\bibinfo {author} {\bibfnamefont {M.}~\bibnamefont
  {Bunney}}, \bibinfo {author} {\bibfnamefont {U.~F.~P.}\ \bibnamefont
  {Seifert}}, \bibinfo {author} {\bibfnamefont {S.}~\bibnamefont {Rachel}},\
  and\ \bibinfo {author} {\bibfnamefont {M.}~\bibnamefont {Vojta}},\ }\bibfield
   {title} {\bibinfo {title} {{Fractionalized Superconductivity Mediated by
  Majorana Fermions in the Kitaev-Kondo Lattice}},\ }\href
  {https://doi.org/10.1103/PhysRevLett.134.206602} {\bibfield  {journal}
  {\bibinfo  {journal} {Phys. Rev. Lett.}\ }\textbf {\bibinfo {volume} {134}},\
  \bibinfo {pages} {206602} (\bibinfo {year} {2025})}\BibitemShut {NoStop}%
\bibitem [{\citenamefont {Murray}\ and\ \citenamefont
  {Vafek}(2014)}]{Murray+14}%
  \BibitemOpen
  \bibfield  {author} {\bibinfo {author} {\bibfnamefont {J.~M.}\ \bibnamefont
  {Murray}}\ and\ \bibinfo {author} {\bibfnamefont {O.}~\bibnamefont {Vafek}},\
  }\bibfield  {title} {\bibinfo {title} {{Renormalization group study of
  interaction-driven quantum anomalous Hall and quantum spin Hall phases in
  quadratic band crossing systems}},\ }\href
  {https://doi.org/10.1103/PhysRevB.89.201110} {\bibfield  {journal} {\bibinfo
  {journal} {Phys. Rev. B}\ }\textbf {\bibinfo {volume} {89}},\ \bibinfo
  {pages} {201110} (\bibinfo {year} {2014})}\BibitemShut {NoStop}%
\bibitem [{\citenamefont {Chubukov}\ \emph {et~al.}(2016)\citenamefont
  {Chubukov}, \citenamefont {Khodas},\ and\ \citenamefont
  {Fernandes}}]{Chubukov+16}%
  \BibitemOpen
  \bibfield  {author} {\bibinfo {author} {\bibfnamefont {A.~V.}\ \bibnamefont
  {Chubukov}}, \bibinfo {author} {\bibfnamefont {M.}~\bibnamefont {Khodas}},\
  and\ \bibinfo {author} {\bibfnamefont {R.~M.}\ \bibnamefont {Fernandes}},\
  }\bibfield  {title} {\bibinfo {title} {{Magnetism, Superconductivity, and
  Spontaneous Orbital Order in Iron-Based Superconductors: Which Comes First
  and Why?}},\ }\href {https://doi.org/10.1103/PhysRevX.6.041045} {\bibfield
  {journal} {\bibinfo  {journal} {Phys. Rev. X}\ }\textbf {\bibinfo {volume}
  {6}},\ \bibinfo {pages} {041045} (\bibinfo {year} {2016})}\BibitemShut
  {NoStop}%
\bibitem [{\citenamefont {Xing}\ \emph {et~al.}(2017)\citenamefont {Xing},
  \citenamefont {Classen}, \citenamefont {Khodas},\ and\ \citenamefont
  {Chubukov}}]{Xing+17}%
  \BibitemOpen
  \bibfield  {author} {\bibinfo {author} {\bibfnamefont {R.-Q.}\ \bibnamefont
  {Xing}}, \bibinfo {author} {\bibfnamefont {L.}~\bibnamefont {Classen}},
  \bibinfo {author} {\bibfnamefont {M.}~\bibnamefont {Khodas}},\ and\ \bibinfo
  {author} {\bibfnamefont {A.~V.}\ \bibnamefont {Chubukov}},\ }\bibfield
  {title} {\bibinfo {title} {{Competing instabilities, orbital ordering, and
  splitting of band degeneracies from a parquet renormalization group analysis
  of a four-pocket model for iron-based superconductors: Application to
  FeSe}},\ }\href {https://doi.org/10.1103/PhysRevB.95.085108} {\bibfield
  {journal} {\bibinfo  {journal} {Phys. Rev. B}\ }\textbf {\bibinfo {volume}
  {95}},\ \bibinfo {pages} {085108} (\bibinfo {year} {2017})}\BibitemShut
  {NoStop}%
\bibitem [{\citenamefont {Parthenios}\ and\ \citenamefont
  {Classen}(2023)}]{Parthenios+23}%
  \BibitemOpen
  \bibfield  {author} {\bibinfo {author} {\bibfnamefont {N.}~\bibnamefont
  {Parthenios}}\ and\ \bibinfo {author} {\bibfnamefont {L.}~\bibnamefont
  {Classen}},\ }\bibfield  {title} {\bibinfo {title} {{Twisted bilayer graphene
  at charge neutrality: Competing orders of SU(4) Dirac fermions}},\ }\href
  {https://doi.org/10.1103/PhysRevB.108.235120} {\bibfield  {journal} {\bibinfo
   {journal} {Phys. Rev. B}\ }\textbf {\bibinfo {volume} {108}},\ \bibinfo
  {pages} {235120} (\bibinfo {year} {2023})}\BibitemShut {NoStop}%
\bibitem [{\citenamefont {Saheli}\ \emph {et~al.}(2024)\citenamefont {Saheli},
  \citenamefont {Lin}, \citenamefont {Hu},\ and\ \citenamefont
  {Kr\"uger}}]{Saheli+24}%
  \BibitemOpen
  \bibfield  {author} {\bibinfo {author} {\bibfnamefont {S.~G.}\ \bibnamefont
  {Saheli}}, \bibinfo {author} {\bibfnamefont {J.}~\bibnamefont {Lin}},
  \bibinfo {author} {\bibfnamefont {H.}~\bibnamefont {Hu}},\ and\ \bibinfo
  {author} {\bibfnamefont {F.}~\bibnamefont {Kr\"uger}},\ }\bibfield  {title}
  {\bibinfo {title} {{Majorana-fermion mean field theories of Kitaev quantum
  spin liquids}},\ }\href {https://doi.org/10.1103/PhysRevB.109.014407}
  {\bibfield  {journal} {\bibinfo  {journal} {Phys. Rev. B}\ }\textbf {\bibinfo
  {volume} {109}},\ \bibinfo {pages} {014407} (\bibinfo {year}
  {2024})}\BibitemShut {NoStop}%
\bibitem [{\citenamefont {Thiagarajan}\ \emph {et~al.}(2024)\citenamefont
  {Thiagarajan}, \citenamefont {Watson}, \citenamefont {Yzeiri}, \citenamefont
  {Hu}, \citenamefont {Uchoa},\ and\ \citenamefont {Kruger}}]{Thiagarajan+24}%
  \BibitemOpen
  \bibfield  {author} {\bibinfo {author} {\bibfnamefont {S.}~\bibnamefont
  {Thiagarajan}}, \bibinfo {author} {\bibfnamefont {C.}~\bibnamefont {Watson}},
  \bibinfo {author} {\bibfnamefont {T.}~\bibnamefont {Yzeiri}}, \bibinfo
  {author} {\bibfnamefont {H.}~\bibnamefont {Hu}}, \bibinfo {author}
  {\bibfnamefont {B.}~\bibnamefont {Uchoa}},\ and\ \bibinfo {author}
  {\bibfnamefont {F.}~\bibnamefont {Kruger}},\ }\bibfield  {title} {\bibinfo
  {title} {{Nature of the Topological Transition of the Kitaev Model in [111]
  Magnetic Field}},\ }\Eprint {https://arxiv.org/abs/2509.13057}
  {arXiv:2509.13057}  (\bibinfo {year} {2024})\BibitemShut {NoStop}%
\bibitem [{\citenamefont {Yadav}\ \emph {et~al.}(2016)\citenamefont {Yadav},
  \citenamefont {Bogdanov}, \citenamefont {Katukuri}, \citenamefont
  {Nishimoto}, \citenamefont {van~den Brink},\ and\ \citenamefont
  {Hozoi}}]{Yadav+16}%
  \BibitemOpen
  \bibfield  {author} {\bibinfo {author} {\bibfnamefont {R.}~\bibnamefont
  {Yadav}}, \bibinfo {author} {\bibfnamefont {N.~A.}\ \bibnamefont {Bogdanov}},
  \bibinfo {author} {\bibfnamefont {V.~M.}\ \bibnamefont {Katukuri}}, \bibinfo
  {author} {\bibfnamefont {S.}~\bibnamefont {Nishimoto}}, \bibinfo {author}
  {\bibfnamefont {J.}~\bibnamefont {van~den Brink}},\ and\ \bibinfo {author}
  {\bibfnamefont {L.}~\bibnamefont {Hozoi}},\ }\bibfield  {title} {\bibinfo
  {title} {{Kitaev exchange and field-induced quantum spin-liquid states in
  honeycomb $\alpha$-RuCl3}},\ }\href {https://doi.org/10.1038/srep37925}
  {\bibfield  {journal} {\bibinfo  {journal} {Scientific Reports}\ }\textbf
  {\bibinfo {volume} {6}},\ \bibinfo {pages} {37925} (\bibinfo {year}
  {2016})}\BibitemShut {NoStop}%
\bibitem [{\citenamefont {Suzuki}\ \emph {et~al.}(2021)\citenamefont {Suzuki},
  \citenamefont {Liu}, \citenamefont {Bertinshaw}, \citenamefont {Ueda},
  \citenamefont {Kim}, \citenamefont {Laha}, \citenamefont {Weber},
  \citenamefont {Yang}, \citenamefont {Wang}, \citenamefont {Takahashi},
  \citenamefont {F{\"u}rsich}, \citenamefont {Minola}, \citenamefont {Lotsch},
  \citenamefont {Kim}, \citenamefont {Yava{\c s}}, \citenamefont {Daghofer},
  \citenamefont {Chaloupka}, \citenamefont {Khaliullin}, \citenamefont
  {Gretarsson},\ and\ \citenamefont {Keimer}}]{Suzuk+21}%
  \BibitemOpen
  \bibfield  {author} {\bibinfo {author} {\bibfnamefont {H.}~\bibnamefont
  {Suzuki}}, \bibinfo {author} {\bibfnamefont {H.}~\bibnamefont {Liu}},
  \bibinfo {author} {\bibfnamefont {J.}~\bibnamefont {Bertinshaw}}, \bibinfo
  {author} {\bibfnamefont {K.}~\bibnamefont {Ueda}}, \bibinfo {author}
  {\bibfnamefont {H.}~\bibnamefont {Kim}}, \bibinfo {author} {\bibfnamefont
  {S.}~\bibnamefont {Laha}}, \bibinfo {author} {\bibfnamefont {D.}~\bibnamefont
  {Weber}}, \bibinfo {author} {\bibfnamefont {Z.}~\bibnamefont {Yang}},
  \bibinfo {author} {\bibfnamefont {L.}~\bibnamefont {Wang}}, \bibinfo {author}
  {\bibfnamefont {H.}~\bibnamefont {Takahashi}}, \bibinfo {author}
  {\bibfnamefont {K.}~\bibnamefont {F{\"u}rsich}}, \bibinfo {author}
  {\bibfnamefont {M.}~\bibnamefont {Minola}}, \bibinfo {author} {\bibfnamefont
  {B.~V.}\ \bibnamefont {Lotsch}}, \bibinfo {author} {\bibfnamefont {B.~J.}\
  \bibnamefont {Kim}}, \bibinfo {author} {\bibfnamefont {H.}~\bibnamefont
  {Yava{\c s}}}, \bibinfo {author} {\bibfnamefont {M.}~\bibnamefont
  {Daghofer}}, \bibinfo {author} {\bibfnamefont {J.}~\bibnamefont {Chaloupka}},
  \bibinfo {author} {\bibfnamefont {G.}~\bibnamefont {Khaliullin}}, \bibinfo
  {author} {\bibfnamefont {H.}~\bibnamefont {Gretarsson}},\ and\ \bibinfo
  {author} {\bibfnamefont {B.}~\bibnamefont {Keimer}},\ }\bibfield  {title}
  {\bibinfo {title} {{Proximate ferromagnetic state in the Kitaev model
  material $\alpha$-RuCl3}},\ }\href
  {https://doi.org/10.1038/s41467-021-24722-4} {\bibfield  {journal} {\bibinfo
  {journal} {Nature Communications}\ }\textbf {\bibinfo {volume} {12}},\
  \bibinfo {pages} {4512} (\bibinfo {year} {2021})}\BibitemShut {NoStop}%
\bibitem [{\citenamefont {Katukuri}\ \emph {et~al.}(2014)\citenamefont
  {Katukuri}, \citenamefont {Nishimoto}, \citenamefont {Yushankhai},
  \citenamefont {Stoyanova}, \citenamefont {Kandpal}, \citenamefont {Choi},
  \citenamefont {Coldea}, \citenamefont {Rousochatzakis}, \citenamefont
  {Hozoi},\ and\ \citenamefont {Brink}}]{Katukuri+14}%
  \BibitemOpen
  \bibfield  {author} {\bibinfo {author} {\bibfnamefont {V.~M.}\ \bibnamefont
  {Katukuri}}, \bibinfo {author} {\bibfnamefont {S.}~\bibnamefont {Nishimoto}},
  \bibinfo {author} {\bibfnamefont {V.}~\bibnamefont {Yushankhai}}, \bibinfo
  {author} {\bibfnamefont {A.}~\bibnamefont {Stoyanova}}, \bibinfo {author}
  {\bibfnamefont {H.}~\bibnamefont {Kandpal}}, \bibinfo {author} {\bibfnamefont
  {S.}~\bibnamefont {Choi}}, \bibinfo {author} {\bibfnamefont {R.}~\bibnamefont
  {Coldea}}, \bibinfo {author} {\bibfnamefont {I.}~\bibnamefont
  {Rousochatzakis}}, \bibinfo {author} {\bibfnamefont {L.}~\bibnamefont
  {Hozoi}},\ and\ \bibinfo {author} {\bibfnamefont {J.~v.~d.}\ \bibnamefont
  {Brink}},\ }\bibfield  {title} {\bibinfo {title} {{Kitaev interactions
  between j = 1/2 moments in honeycomb Na2IrO3 are large and ferromagnetic:
  insights from ab initio quantum chemistry calculations}},\ }\href
  {https://doi.org/10.1088/1367-2630/16/1/013056} {\bibfield  {journal}
  {\bibinfo  {journal} {New Journal of Physics}\ }\textbf {\bibinfo {volume}
  {16}},\ \bibinfo {pages} {013056} (\bibinfo {year} {2014})}\BibitemShut
  {NoStop}%
\bibitem [{\citenamefont {Songvilay}\ \emph {et~al.}(2020)\citenamefont
  {Songvilay}, \citenamefont {Robert}, \citenamefont {Petit}, \citenamefont
  {Rodriguez-Rivera}, \citenamefont {Ratcliff}, \citenamefont {Damay},
  \citenamefont {Bal\'edent}, \citenamefont {Jim\'enez-Ruiz}, \citenamefont
  {Lejay}, \citenamefont {Pachoud}, \citenamefont {Hadj-Azzem}, \citenamefont
  {Simonet},\ and\ \citenamefont {Stock}}]{Songvilay+20}%
  \BibitemOpen
  \bibfield  {author} {\bibinfo {author} {\bibfnamefont {M.}~\bibnamefont
  {Songvilay}}, \bibinfo {author} {\bibfnamefont {J.}~\bibnamefont {Robert}},
  \bibinfo {author} {\bibfnamefont {S.}~\bibnamefont {Petit}}, \bibinfo
  {author} {\bibfnamefont {J.~A.}\ \bibnamefont {Rodriguez-Rivera}}, \bibinfo
  {author} {\bibfnamefont {W.~D.}\ \bibnamefont {Ratcliff}}, \bibinfo {author}
  {\bibfnamefont {F.}~\bibnamefont {Damay}}, \bibinfo {author} {\bibfnamefont
  {V.}~\bibnamefont {Bal\'edent}}, \bibinfo {author} {\bibfnamefont
  {M.}~\bibnamefont {Jim\'enez-Ruiz}}, \bibinfo {author} {\bibfnamefont
  {P.}~\bibnamefont {Lejay}}, \bibinfo {author} {\bibfnamefont
  {E.}~\bibnamefont {Pachoud}}, \bibinfo {author} {\bibfnamefont
  {A.}~\bibnamefont {Hadj-Azzem}}, \bibinfo {author} {\bibfnamefont
  {V.}~\bibnamefont {Simonet}},\ and\ \bibinfo {author} {\bibfnamefont
  {C.}~\bibnamefont {Stock}},\ }\bibfield  {title} {\bibinfo {title} {{Kitaev
  interactions in the Co honeycomb antiferromagnets
  ${\mathrm{Na}}_{3}{\mathrm{Co}}_{2}{\mathrm{SbO}}_{6}$ and
  ${\mathrm{Na}}_{2}{\mathrm{Co}}_{2}{\mathrm{TeO}}_{6}$}},\ }\href
  {https://doi.org/10.1103/PhysRevB.102.224429} {\bibfield  {journal} {\bibinfo
   {journal} {Phys. Rev. B}\ }\textbf {\bibinfo {volume} {102}},\ \bibinfo
  {pages} {224429} (\bibinfo {year} {2020})}\BibitemShut {NoStop}%
\bibitem [{\citenamefont {Castro~Neto}\ \emph {et~al.}(2009)\citenamefont
  {Castro~Neto}, \citenamefont {Guinea}, \citenamefont {Peres}, \citenamefont
  {Novoselov},\ and\ \citenamefont {Geim}}]{Neto+09}%
  \BibitemOpen
  \bibfield  {author} {\bibinfo {author} {\bibfnamefont {A.~H.}\ \bibnamefont
  {Castro~Neto}}, \bibinfo {author} {\bibfnamefont {F.}~\bibnamefont {Guinea}},
  \bibinfo {author} {\bibfnamefont {N.~M.~R.}\ \bibnamefont {Peres}}, \bibinfo
  {author} {\bibfnamefont {K.~S.}\ \bibnamefont {Novoselov}},\ and\ \bibinfo
  {author} {\bibfnamefont {A.~K.}\ \bibnamefont {Geim}},\ }\bibfield  {title}
  {\bibinfo {title} {{The electronic properties of graphene}},\ }\href
  {https://doi.org/10.1103/RevModPhys.81.109} {\bibfield  {journal} {\bibinfo
  {journal} {Rev. Mod. Phys.}\ }\textbf {\bibinfo {volume} {81}},\ \bibinfo
  {pages} {109} (\bibinfo {year} {2009})}\BibitemShut {NoStop}%
\bibitem [{\citenamefont {Herbut}(2006)}]{Herbut06}%
  \BibitemOpen
  \bibfield  {author} {\bibinfo {author} {\bibfnamefont {I.~F.}\ \bibnamefont
  {Herbut}},\ }\bibfield  {title} {\bibinfo {title} {{Interactions and Phase
  Transitions on Graphene's Honeycomb Lattice}},\ }\href
  {https://doi.org/10.1103/PhysRevLett.97.146401} {\bibfield  {journal}
  {\bibinfo  {journal} {Phys. Rev. Lett.}\ }\textbf {\bibinfo {volume} {97}},\
  \bibinfo {pages} {146401} (\bibinfo {year} {2006})}\BibitemShut {NoStop}%
\bibitem [{\citenamefont {Herbut}\ \emph {et~al.}(2009)\citenamefont {Herbut},
  \citenamefont {Juricic},\ and\ \citenamefont {Roy}}]{Herbut+09}%
  \BibitemOpen
  \bibfield  {author} {\bibinfo {author} {\bibfnamefont {I.~F.}\ \bibnamefont
  {Herbut}}, \bibinfo {author} {\bibfnamefont {V.}~\bibnamefont {Juricic}},\
  and\ \bibinfo {author} {\bibfnamefont {B.}~\bibnamefont {Roy}},\ }\bibfield
  {title} {\bibinfo {title} {{Theory of interacting electrons on the honeycomb
  lattice}},\ }\href {https://doi.org/10.1103/PhysRevB.79.085116} {\bibfield
  {journal} {\bibinfo  {journal} {Phys. Rev. B}\ }\textbf {\bibinfo {volume}
  {79}},\ \bibinfo {pages} {085116} (\bibinfo {year} {2009})}\BibitemShut
  {NoStop}%
\bibitem [{\citenamefont {Cardy}(1996)}]{Cardy_1996}%
  \BibitemOpen
  \bibfield  {author} {\bibinfo {author} {\bibfnamefont {J.}~\bibnamefont
  {Cardy}},\ }\href@noop {} {\emph {\bibinfo {title} {{Scaling and
  Renormalization in Statistical Physics}}}},\ Cambridge Lecture Notes in
  Physics\ (\bibinfo  {publisher} {Cambridge University Press},\ \bibinfo
  {year} {1996})\BibitemShut {NoStop}%
\bibitem [{\citenamefont {Ladovrechis}\ \emph {et~al.}(2023)\citenamefont
  {Ladovrechis}, \citenamefont {Ray}, \citenamefont {Meng},\ and\ \citenamefont
  {Janssen}}]{Ladovrechis+23}%
  \BibitemOpen
  \bibfield  {author} {\bibinfo {author} {\bibfnamefont {K.}~\bibnamefont
  {Ladovrechis}}, \bibinfo {author} {\bibfnamefont {S.}~\bibnamefont {Ray}},
  \bibinfo {author} {\bibfnamefont {T.}~\bibnamefont {Meng}},\ and\ \bibinfo
  {author} {\bibfnamefont {L.}~\bibnamefont {Janssen}},\ }\bibfield  {title}
  {\bibinfo {title} {{Gross-Neveu-Heisenberg criticality from
  $2+\ensuremath{\epsilon}$ expansion}},\ }\href
  {https://doi.org/10.1103/PhysRevB.107.035151} {\bibfield  {journal} {\bibinfo
   {journal} {Phys. Rev. B}\ }\textbf {\bibinfo {volume} {107}},\ \bibinfo
  {pages} {035151} (\bibinfo {year} {2023})}\BibitemShut {NoStop}%
\bibitem [{\citenamefont {Zerf}\ \emph {et~al.}(2017)\citenamefont {Zerf},
  \citenamefont {Mihaila}, \citenamefont {Marquard}, \citenamefont {Herbut},\
  and\ \citenamefont {Scherer}}]{Zerf+17}%
  \BibitemOpen
  \bibfield  {author} {\bibinfo {author} {\bibfnamefont {N.}~\bibnamefont
  {Zerf}}, \bibinfo {author} {\bibfnamefont {L.~N.}\ \bibnamefont {Mihaila}},
  \bibinfo {author} {\bibfnamefont {P.}~\bibnamefont {Marquard}}, \bibinfo
  {author} {\bibfnamefont {I.~F.}\ \bibnamefont {Herbut}},\ and\ \bibinfo
  {author} {\bibfnamefont {M.~M.}\ \bibnamefont {Scherer}},\ }\bibfield
  {title} {\bibinfo {title} {{Four-loop critical exponents for the
  Gross-Neveu-Yukawa models}},\ }\href
  {https://doi.org/10.1103/PhysRevD.96.096010} {\bibfield  {journal} {\bibinfo
  {journal} {Phys. Rev. D}\ }\textbf {\bibinfo {volume} {96}},\ \bibinfo
  {pages} {096010} (\bibinfo {year} {2017})}\BibitemShut {NoStop}%
\bibitem [{\citenamefont {Gracey}(2018)}]{Gracey18}%
  \BibitemOpen
  \bibfield  {author} {\bibinfo {author} {\bibfnamefont {J.~A.}\ \bibnamefont
  {Gracey}},\ }\bibfield  {title} {\bibinfo {title} {{Large $N$ critical
  exponents for the chiral Heisenberg Gross-Neveu universality class}},\ }\href
  {https://doi.org/10.1103/PhysRevD.97.105009} {\bibfield  {journal} {\bibinfo
  {journal} {Phys. Rev. D}\ }\textbf {\bibinfo {volume} {97}},\ \bibinfo
  {pages} {105009} (\bibinfo {year} {2018})}\BibitemShut {NoStop}%
\bibitem [{\citenamefont {Janssen}\ and\ \citenamefont
  {Herbut}(2014)}]{Janssen+14}%
  \BibitemOpen
  \bibfield  {author} {\bibinfo {author} {\bibfnamefont {L.}~\bibnamefont
  {Janssen}}\ and\ \bibinfo {author} {\bibfnamefont {I.~F.}\ \bibnamefont
  {Herbut}},\ }\bibfield  {title} {\bibinfo {title} {{Antiferromagnetic
  critical point on graphene's honeycomb lattice: A functional renormalization
  group approach}},\ }\href {https://doi.org/10.1103/PhysRevB.89.205403}
  {\bibfield  {journal} {\bibinfo  {journal} {Phys. Rev. B}\ }\textbf {\bibinfo
  {volume} {89}},\ \bibinfo {pages} {205403} (\bibinfo {year}
  {2014})}\BibitemShut {NoStop}%
\bibitem [{\citenamefont {Knorr}(2018)}]{Knorr18}%
  \BibitemOpen
  \bibfield  {author} {\bibinfo {author} {\bibfnamefont {B.}~\bibnamefont
  {Knorr}},\ }\bibfield  {title} {\bibinfo {title} {{Critical chiral Heisenberg
  model with the functional renormalization group}},\ }\href
  {https://doi.org/10.1103/PhysRevB.97.075129} {\bibfield  {journal} {\bibinfo
  {journal} {Phys. Rev. B}\ }\textbf {\bibinfo {volume} {97}},\ \bibinfo
  {pages} {075129} (\bibinfo {year} {2018})}\BibitemShut {NoStop}%
\bibitem [{\citenamefont {Parisen~Toldin}\ \emph {et~al.}(2015)\citenamefont
  {Parisen~Toldin}, \citenamefont {Hohenadler}, \citenamefont {Assaad},\ and\
  \citenamefont {Herbut}}]{Parisen+15}%
  \BibitemOpen
  \bibfield  {author} {\bibinfo {author} {\bibfnamefont {F.}~\bibnamefont
  {Parisen~Toldin}}, \bibinfo {author} {\bibfnamefont {M.}~\bibnamefont
  {Hohenadler}}, \bibinfo {author} {\bibfnamefont {F.~F.}\ \bibnamefont
  {Assaad}},\ and\ \bibinfo {author} {\bibfnamefont {I.~F.}\ \bibnamefont
  {Herbut}},\ }\bibfield  {title} {\bibinfo {title} {{Fermionic quantum
  criticality in honeycomb and $\ensuremath{\pi}$-flux Hubbard models:
  Finite-size scaling of renormalization-group-invariant observables from
  quantum Monte Carlo}},\ }\href {https://doi.org/10.1103/PhysRevB.91.165108}
  {\bibfield  {journal} {\bibinfo  {journal} {Phys. Rev. B}\ }\textbf {\bibinfo
  {volume} {91}},\ \bibinfo {pages} {165108} (\bibinfo {year}
  {2015})}\BibitemShut {NoStop}%
\bibitem [{\citenamefont {Liu}\ \emph {et~al.}(2019)\citenamefont {Liu},
  \citenamefont {Wang}, \citenamefont {Sato}, \citenamefont {Hohenadler},
  \citenamefont {Wang}, \citenamefont {Guo},\ and\ \citenamefont
  {Assaad}}]{Liu+19}%
  \BibitemOpen
  \bibfield  {author} {\bibinfo {author} {\bibfnamefont {Y.}~\bibnamefont
  {Liu}}, \bibinfo {author} {\bibfnamefont {Z.}~\bibnamefont {Wang}}, \bibinfo
  {author} {\bibfnamefont {T.}~\bibnamefont {Sato}}, \bibinfo {author}
  {\bibfnamefont {M.}~\bibnamefont {Hohenadler}}, \bibinfo {author}
  {\bibfnamefont {C.}~\bibnamefont {Wang}}, \bibinfo {author} {\bibfnamefont
  {W.}~\bibnamefont {Guo}},\ and\ \bibinfo {author} {\bibfnamefont {F.~F.}\
  \bibnamefont {Assaad}},\ }\bibfield  {title} {\bibinfo {title}
  {{Superconductivity from the condensation of topological defects in a quantum
  spin-Hall insulator}},\ }\href {https://doi.org/10.1038/s41467-019-10372-0}
  {\bibfield  {journal} {\bibinfo  {journal} {Nature Communications}\ }\textbf
  {\bibinfo {volume} {10}},\ \bibinfo {pages} {2658} (\bibinfo {year}
  {2019})}\BibitemShut {NoStop}%
\bibitem [{\citenamefont {Liu}\ \emph {et~al.}(2021)\citenamefont {Liu},
  \citenamefont {Wang}, \citenamefont {Sato}, \citenamefont {Guo},\ and\
  \citenamefont {Assaad}}]{Liu+21}%
  \BibitemOpen
  \bibfield  {author} {\bibinfo {author} {\bibfnamefont {Y.}~\bibnamefont
  {Liu}}, \bibinfo {author} {\bibfnamefont {Z.}~\bibnamefont {Wang}}, \bibinfo
  {author} {\bibfnamefont {T.}~\bibnamefont {Sato}}, \bibinfo {author}
  {\bibfnamefont {W.}~\bibnamefont {Guo}},\ and\ \bibinfo {author}
  {\bibfnamefont {F.~F.}\ \bibnamefont {Assaad}},\ }\bibfield  {title}
  {\bibinfo {title} {{Gross-Neveu Heisenberg criticality: Dynamical generation
  of quantum spin Hall masses}},\ }\href
  {https://doi.org/10.1103/PhysRevB.104.035107} {\bibfield  {journal} {\bibinfo
   {journal} {Phys. Rev. B}\ }\textbf {\bibinfo {volume} {104}},\ \bibinfo
  {pages} {035107} (\bibinfo {year} {2021})}\BibitemShut {NoStop}%
\bibitem [{\citenamefont {Otsuka}\ \emph {et~al.}(2016)\citenamefont {Otsuka},
  \citenamefont {Yunoki},\ and\ \citenamefont {Sorella}}]{Otsuka+16}%
  \BibitemOpen
  \bibfield  {author} {\bibinfo {author} {\bibfnamefont {Y.}~\bibnamefont
  {Otsuka}}, \bibinfo {author} {\bibfnamefont {S.}~\bibnamefont {Yunoki}},\
  and\ \bibinfo {author} {\bibfnamefont {S.}~\bibnamefont {Sorella}},\
  }\bibfield  {title} {\bibinfo {title} {{Universal Quantum Criticality in the
  Metal-Insulator Transition of Two-Dimensional Interacting Dirac Electrons}},\
  }\href {https://doi.org/10.1103/PhysRevX.6.011029} {\bibfield  {journal}
  {\bibinfo  {journal} {Phys. Rev. X}\ }\textbf {\bibinfo {volume} {6}},\
  \bibinfo {pages} {011029} (\bibinfo {year} {2016})}\BibitemShut {NoStop}%
\bibitem [{\citenamefont {Otsuka}\ \emph {et~al.}(2020)\citenamefont {Otsuka},
  \citenamefont {Seki}, \citenamefont {Sorella},\ and\ \citenamefont
  {Yunoki}}]{Otsuka+20}%
  \BibitemOpen
  \bibfield  {author} {\bibinfo {author} {\bibfnamefont {Y.}~\bibnamefont
  {Otsuka}}, \bibinfo {author} {\bibfnamefont {K.}~\bibnamefont {Seki}},
  \bibinfo {author} {\bibfnamefont {S.}~\bibnamefont {Sorella}},\ and\ \bibinfo
  {author} {\bibfnamefont {S.}~\bibnamefont {Yunoki}},\ }\bibfield  {title}
  {\bibinfo {title} {{Dirac electrons in the square-lattice Hubbard model with
  a $d$-wave pairing field: The chiral Heisenberg universality class
  revisited}},\ }\href {https://doi.org/10.1103/PhysRevB.102.235105} {\bibfield
   {journal} {\bibinfo  {journal} {Phys. Rev. B}\ }\textbf {\bibinfo {volume}
  {102}},\ \bibinfo {pages} {235105} (\bibinfo {year} {2020})}\BibitemShut
  {NoStop}%
\bibitem [{\citenamefont {Xu}\ and\ \citenamefont {Grover}(2021)}]{Xu+21}%
  \BibitemOpen
  \bibfield  {author} {\bibinfo {author} {\bibfnamefont {X.~Y.}\ \bibnamefont
  {Xu}}\ and\ \bibinfo {author} {\bibfnamefont {T.}~\bibnamefont {Grover}},\
  }\bibfield  {title} {\bibinfo {title} {{Competing Nodal $d$-Wave
  Superconductivity and Antiferromagnetism}},\ }\href
  {https://doi.org/10.1103/PhysRevLett.126.217002} {\bibfield  {journal}
  {\bibinfo  {journal} {Phys. Rev. Lett.}\ }\textbf {\bibinfo {volume} {126}},\
  \bibinfo {pages} {217002} (\bibinfo {year} {2021})}\BibitemShut {NoStop}%
\bibitem [{\citenamefont {Buividovich}\ \emph {et~al.}(2018)\citenamefont
  {Buividovich}, \citenamefont {Smith}, \citenamefont {Ulybyshev},\ and\
  \citenamefont {von Smekal}}]{Buividovich+18}%
  \BibitemOpen
  \bibfield  {author} {\bibinfo {author} {\bibfnamefont {P.}~\bibnamefont
  {Buividovich}}, \bibinfo {author} {\bibfnamefont {D.}~\bibnamefont {Smith}},
  \bibinfo {author} {\bibfnamefont {M.}~\bibnamefont {Ulybyshev}},\ and\
  \bibinfo {author} {\bibfnamefont {L.}~\bibnamefont {von Smekal}},\ }\bibfield
   {title} {\bibinfo {title} {{Hybrid Monte Carlo study of competing order in
  the extended fermionic Hubbard model on the hexagonal lattice}},\ }\href
  {https://doi.org/10.1103/PhysRevB.98.235129} {\bibfield  {journal} {\bibinfo
  {journal} {Phys. Rev. B}\ }\textbf {\bibinfo {volume} {98}},\ \bibinfo
  {pages} {235129} (\bibinfo {year} {2018})}\BibitemShut {NoStop}%
\bibitem [{\citenamefont {Buividovich}\ \emph {et~al.}(2019)\citenamefont
  {Buividovich}, \citenamefont {Smith}, \citenamefont {Ulybyshev},\ and\
  \citenamefont {von Smekal}}]{Buividovich+19}%
  \BibitemOpen
  \bibfield  {author} {\bibinfo {author} {\bibfnamefont {P.}~\bibnamefont
  {Buividovich}}, \bibinfo {author} {\bibfnamefont {D.}~\bibnamefont {Smith}},
  \bibinfo {author} {\bibfnamefont {M.}~\bibnamefont {Ulybyshev}},\ and\
  \bibinfo {author} {\bibfnamefont {L.}~\bibnamefont {von Smekal}},\ }\bibfield
   {title} {\bibinfo {title} {{Numerical evidence of conformal phase transition
  in graphene with long-range interactions}},\ }\href
  {https://doi.org/10.1103/PhysRevB.99.205434} {\bibfield  {journal} {\bibinfo
  {journal} {Phys. Rev. B}\ }\textbf {\bibinfo {volume} {99}},\ \bibinfo
  {pages} {205434} (\bibinfo {year} {2019})}\BibitemShut {NoStop}%
\bibitem [{\citenamefont {Ostmeyer}\ \emph {et~al.}(2020)\citenamefont
  {Ostmeyer}, \citenamefont {Berkowitz}, \citenamefont {Krieg}, \citenamefont
  {L\"ahde}, \citenamefont {Luu},\ and\ \citenamefont {Urbach}}]{Ostmeyer+20}%
  \BibitemOpen
  \bibfield  {author} {\bibinfo {author} {\bibfnamefont {J.}~\bibnamefont
  {Ostmeyer}}, \bibinfo {author} {\bibfnamefont {E.}~\bibnamefont {Berkowitz}},
  \bibinfo {author} {\bibfnamefont {S.}~\bibnamefont {Krieg}}, \bibinfo
  {author} {\bibfnamefont {T.~A.}\ \bibnamefont {L\"ahde}}, \bibinfo {author}
  {\bibfnamefont {T.}~\bibnamefont {Luu}},\ and\ \bibinfo {author}
  {\bibfnamefont {C.}~\bibnamefont {Urbach}},\ }\bibfield  {title} {\bibinfo
  {title} {{Semimetal--Mott insulator quantum phase transition of the Hubbard
  model on the honeycomb lattice}},\ }\href
  {https://doi.org/10.1103/PhysRevB.102.245105} {\bibfield  {journal} {\bibinfo
   {journal} {Phys. Rev. B}\ }\textbf {\bibinfo {volume} {102}},\ \bibinfo
  {pages} {245105} (\bibinfo {year} {2020})}\BibitemShut {NoStop}%
\bibitem [{\citenamefont {Ostmeyer}\ \emph {et~al.}(2021)\citenamefont
  {Ostmeyer}, \citenamefont {Berkowitz}, \citenamefont {Krieg}, \citenamefont
  {L\"ahde}, \citenamefont {Luu},\ and\ \citenamefont {Urbach}}]{Ostmeyer+21}%
  \BibitemOpen
  \bibfield  {author} {\bibinfo {author} {\bibfnamefont {J.}~\bibnamefont
  {Ostmeyer}}, \bibinfo {author} {\bibfnamefont {E.}~\bibnamefont {Berkowitz}},
  \bibinfo {author} {\bibfnamefont {S.}~\bibnamefont {Krieg}}, \bibinfo
  {author} {\bibfnamefont {T.~A.}\ \bibnamefont {L\"ahde}}, \bibinfo {author}
  {\bibfnamefont {T.}~\bibnamefont {Luu}},\ and\ \bibinfo {author}
  {\bibfnamefont {C.}~\bibnamefont {Urbach}},\ }\bibfield  {title} {\bibinfo
  {title} {{Antiferromagnetic character of the quantum phase transition in the
  Hubbard model on the honeycomb lattice}},\ }\href
  {https://doi.org/10.1103/PhysRevB.104.155142} {\bibfield  {journal} {\bibinfo
   {journal} {Phys. Rev. B}\ }\textbf {\bibinfo {volume} {104}},\ \bibinfo
  {pages} {155142} (\bibinfo {year} {2021})}\BibitemShut {NoStop}%
\bibitem [{\citenamefont {Li}\ \emph {et~al.}(2017)\citenamefont {Li},
  \citenamefont {Jiang}, \citenamefont {Jian},\ and\ \citenamefont
  {Yao}}]{Li+17}%
  \BibitemOpen
  \bibfield  {author} {\bibinfo {author} {\bibfnamefont {Z.-X.}\ \bibnamefont
  {Li}}, \bibinfo {author} {\bibfnamefont {Y.-F.}\ \bibnamefont {Jiang}},
  \bibinfo {author} {\bibfnamefont {S.-K.}\ \bibnamefont {Jian}},\ and\
  \bibinfo {author} {\bibfnamefont {H.}~\bibnamefont {Yao}},\ }\bibfield
  {title} {\bibinfo {title} {{Fermion-induced quantum critical points}},\
  }\href {https://doi.org/10.1038/s41467-017-00167-6} {\bibfield  {journal}
  {\bibinfo  {journal} {Nature Communications}\ }\textbf {\bibinfo {volume}
  {8}},\ \bibinfo {pages} {314} (\bibinfo {year} {2017})}\BibitemShut {NoStop}%
\bibitem [{\citenamefont {Christou}\ \emph {et~al.}(2020)\citenamefont
  {Christou}, \citenamefont {de~Juan},\ and\ \citenamefont
  {Kr\"uger}}]{Christou+20}%
  \BibitemOpen
  \bibfield  {author} {\bibinfo {author} {\bibfnamefont {E.}~\bibnamefont
  {Christou}}, \bibinfo {author} {\bibfnamefont {F.}~\bibnamefont {de~Juan}},\
  and\ \bibinfo {author} {\bibfnamefont {F.}~\bibnamefont {Kr\"uger}},\
  }\bibfield  {title} {\bibinfo {title} {{Criticality of Dirac fermions in the
  presence of emergent gauge fields}},\ }\href
  {https://doi.org/10.1103/PhysRevB.101.155121} {\bibfield  {journal} {\bibinfo
   {journal} {Phys. Rev. B}\ }\textbf {\bibinfo {volume} {101}},\ \bibinfo
  {pages} {155121} (\bibinfo {year} {2020})}\BibitemShut {NoStop}%
\bibitem [{\citenamefont {Tang}\ and\ \citenamefont {Yao}(2025)}]{Tang+25}%
  \BibitemOpen
  \bibfield  {author} {\bibinfo {author} {\bibfnamefont {C.}~\bibnamefont
  {Tang}}\ and\ \bibinfo {author} {\bibfnamefont {H.}~\bibnamefont {Yao}},\
  }\href {https://arxiv.org/abs/2512.17729} {\bibinfo {title} {{Fractionalized
  topological d+id superconductivity in the Yao-Lee-Kondo model}}} (\bibinfo
  {year} {2025}),\ \Eprint {https://arxiv.org/abs/2512.17729} {arXiv:2512.17729
  [cond-mat.str-el]} \BibitemShut {NoStop}%
\bibitem [{\citenamefont {Yao}\ and\ \citenamefont {Lee}(2011)}]{Yao+11}%
  \BibitemOpen
  \bibfield  {author} {\bibinfo {author} {\bibfnamefont {H.}~\bibnamefont
  {Yao}}\ and\ \bibinfo {author} {\bibfnamefont {D.-H.}\ \bibnamefont {Lee}},\
  }\bibfield  {title} {\bibinfo {title} {{Fermionic Magnons, Non-Abelian
  Spinons, and the Spin Quantum Hall Effect from an Exactly Solvable Spin-$1/2$
  Kitaev Model with SU(2) Symmetry}},\ }\href
  {https://doi.org/10.1103/PhysRevLett.107.087205} {\bibfield  {journal}
  {\bibinfo  {journal} {Phys. Rev. Lett.}\ }\textbf {\bibinfo {volume} {107}},\
  \bibinfo {pages} {087205} (\bibinfo {year} {2011})}\BibitemShut {NoStop}%
\bibitem [{\citenamefont {Balents}\ \emph {et~al.}(2002)\citenamefont
  {Balents}, \citenamefont {Fisher},\ and\ \citenamefont
  {Girvin}}]{Balents+02}%
  \BibitemOpen
  \bibfield  {author} {\bibinfo {author} {\bibfnamefont {L.}~\bibnamefont
  {Balents}}, \bibinfo {author} {\bibfnamefont {M.~P.~A.}\ \bibnamefont
  {Fisher}},\ and\ \bibinfo {author} {\bibfnamefont {S.~M.}\ \bibnamefont
  {Girvin}},\ }\bibfield  {title} {\bibinfo {title} {{Fractionalization in an
  easy-axis Kagome antiferromagnet}},\ }\href
  {https://doi.org/10.1103/PhysRevB.65.224412} {\bibfield  {journal} {\bibinfo
  {journal} {Phys. Rev. B}\ }\textbf {\bibinfo {volume} {65}},\ \bibinfo
  {pages} {224412} (\bibinfo {year} {2002})}\BibitemShut {NoStop}%
\end{thebibliography}
\end{document}